\documentclass{aa}
\usepackage{graphicx}
\usepackage[varg]{txfonts}
\usepackage{natbib}

\def\tex {\ifmmode{{T}_{\rm ex}}\else{$T_{\rm ex}$}\fi}
\def\tmb {\ifmmode{{T}_{\rm mb}}\else{$T_{\rm mb}$}\fi}
\def\ci     {\ifmmode{{\rm C}{\rm \small I}}\else{C\ts {\scriptsize I}}\fi}
\def\hi     {\ifmmode{{\rm H}{\rm \small I}}\else{H\ts {\scriptsize I}}\fi}
\def\hh     {\ifmmode{{\rm H}_2}\else{H$_2$}\fi}

\def\ts     {\thinspace}
\def\kms    {\ifmmode{{\rm \ts km\ts s}^{-1}}\else{\ts km\ts s$^{-1}$}\fi}
\def\msol   {\ifmmode{{\rm M}_{\odot}}\else{M$_{\odot}$}\fi}
\def\lsol   {\ifmmode{{\rm L}_{\odot}}\else{L$_{\odot}$}\fi}
\def\zsol   {\ifmmode{{\rm Z}_{\odot}}\else{Z$_{\odot}$}\fi}

\begin{document}

\title{The molecular gas content of shell galaxies
\thanks{Based on observations carried out with the 30m telescope
of IRAM (Institute of RAdioastronomy in Millimeter).
}}

\author{B. Mancillas \inst{1}
\and
F. Combes \inst{1,2}
\and
P.-A. Duc \inst{3}
           }
\institute{Observatoire de Paris, LERMA, PSL University,
CNRS, Sorbonne University, UPMC, Paris, France
  \and 
  Coll\`ege de France, 11 Place Marcelin Berthelot, 75005 Paris
  \and 
  Universit\'e de Strasbourg, CNRS, Observatoire de Strasbourg, F-67000 Strasbourg, France
  }

   \date{Received  2019/ Accepted  2019}

   \titlerunning{Molecular gas content in shell galaxies}
   \authorrunning{Mancillas et al.}

   \abstract{Shells are fine stellar structures identified by their arc-like shapes present 
   around a galaxy and currently thought to be vestiges of galaxy interactions and/or mergers. 
   The study of their number, geometry, stellar populations and gas content can help
   to derive the interaction/merger history of a galaxy.
    Numerical simulations have proposed
   a mechanism of shell formation through phase wrapping during a radial minor merger.
   Alternatively, there could be merely a space wrapping, when particles have not made
   any radial oscillation yet, but are bound by their radial expansion, or produce
   an edge-brightened feature. These can be distinguished, because they are expected
   to keep a high radial velocity.
   While shells are first a stellar phenomenon, HI and CO observations have revealed 
   neutral gas associated with shells. Some of the gas, the most diffuse and 
   dissipative, is expected to be 
   driven quickly to the center if it is travelling on 
   nearly radial orbits.  Molecular gas, distributed in
   dense clumps, is less dissipative, and may be associated to shells.  It can then determine
   the shell velocity, too difficult to obtain from stars. 
   We present here a search for molecular gas in nine shell galaxies with the 
   IRAM-30m telescope. Six of them are detected in their galaxy center, 
   and in three galaxies, we clearly detect molecular gas in shells. The derived 
   amount of molecular
   gas varies from 1.5 10$^8$ to 3.4 10$^9$ M$_\odot$ in the shells. For two of
   them (Arp~10 and NGC~3656), the shells are characteristic of an oblate system. Their
   velocity is nearly systemic, and we conclude that these shells are phase-wrapped.
   For the third one (NGC~3934) the shells appear to participate in the rotation.
     Follow up with higher spatial resolution is required to conclude.
}
     
\maketitle


\section{Introduction}
It is well accepted that galaxy interactions frequently lead to merging, and must have played
a fundamental role in the formation and evolution of galaxies \citep{Toomre1972}. Most of our
understanding of galaxy interactions has been obtained through numerical simulations 
\citep[see for a review][]{Barnes1992}. Among the best vestiges of galaxy interactions are shells and 
ripples \citep{Schweizer1992}, which are often present around merger remnants. Shells, which
are composed by stars, were originally observed around elliptical and S0 galaxies which are 
poor-gas systems \citep{Carter1982}. A convincing theory has been developed to account for the 
formation of shells around early-type galaxies as a {\it{phase-wrapping phenomenon}} 
\citep{Quinn1984,Dupraz1986,Dupraz1987,Hernquist1989}. According to the theory, the stars modeled 
as collisionless particles, which were originally bound to the merging companions, are liberated 
and oscillate with different periods of oscillation, around the primary galaxy.
They accumulate near the apocenters of their 
orbits, to form shell-like features. This phenomenon is also predicted to occur in the last phase of
any merging between two {\it{spiral}} galaxies. The end-product appears to be an elliptical, with 
a de Vaucouleurs law as light profile \citep[e.g.][]{Schweizer1990}. The late infall of tidal tails and debris,
predominantly consisting of stars, initiates the phase-wrapping and subsequent shell formation. 
Most spiral merger remnants, such as the prototypical NGC 7252, appear to indeed possess 
shells \citep{Schweizer1990}.

The morphology of shells may help to reveal the 3D-structure of the galaxy potential:
if the galaxy is prolate, shells align along the major axis, and are interleaved in radius,
while shells wind up randomly in azimuth near the equatorial plane of an oblate galaxy
\citep{Dupraz1986}. All these are phase-wrapped shells, where stars are accumulating
at the apocenter of their oscillations. However, as \citet{Hernquist1988} have pointed out, there
could also be some features like tails or streams, edge-brightened or only space-bounded, 
which do not fall under phase wrapping, but only under space-wrapping.

This space-wrapping scenario was invoked when neutral hydrogen was observed in 
systems containing shells. \cite{Schiminovich1994} and \cite{Schiminovich1995} 
have detected diffuse HI gas 
associated with the stellar shells (Centaurus A and NGC 2865). 
These observations were a priori surprising, in the phase-wrapped scenario, since
we believe that the diffuse gaseous and stellar components do not have the same behavior 
when approaching to the center of the potential well in quasi-radial orbits \citep{Weil1993}. 
 The gas was expected to infall to the center, without being able to oscillate back.
 However, part of the gas could be structured in clouds and less dissipative, and also
 the orbits might be less radial. Simulating a minor merger with clumpy gas, \citet{Combes2000} have 
 shown that indeed, a large fraction of the gas could follow the stars in phase-wrapped shells.
  Since the gas in the outer parts of the merging companion is less bound than the stars, it is 
  liberated first, and less slowed down by dynamical friction: gaseous shells have a slightly 
  larger radius than the stellar ones. In Centaurus A, the HI is associated 
with the outermost faint stellar shells. It is displaced to the outside of the optical shells, and has 
a high rotation velocity, suggesting that only a companion with a more circular orbit is able to 
produce this HI ring \citep{Schiminovich1994}. Other cases of HI shells 
displaced with respect to the stellar shells have been reported \citep{Schiminovich1997,vanGorkom2000}.

Detecting gas in shells is useful to be able to determine their velocity. Indeed,
getting stellar velocity is costly and requires the use of spectrographs with IFU capabilities
on large telescopes. The HI maps suffer from low
spatial resolution, so molecular gas might be one of the best ways to determine
shell kinematics.  The latter is important to try discriminating between the
different scenarios: in the phase-wrapping model, stars have zero radial
velocity in the shells, which correspond to their apocenter. Velocities might
be much higher in the space-wrapping model.
In order to get more insight into the process of gaseous shell formation, and
into the behavior of gas in shell galaxies in general, 
 we performed observations of CO(1-0) and CO(2-1) emission
lines in the shell regions of eight early-type 
galaxies, two of them taken from the CO survey of $ATLAS^{3D}$ sample 
\citep{Young2011}, and five of them have H$_{2}$ gas 
 detected in their center.
 In all cases, we re-observed the center positions, to obtain better S/N ratio.
We detected CO emission associated with shells in three galaxies and 
detected molecular gas in six galaxies, where we made small maps.
Out of the 9 systems observed only two 
were not detected at all (NGC~5018 and NGC~7600).
 NGC~474 also is not detected in its center, but CO is present in
  its companion.
 We then examine the expected association between the 
stellar and gaseous (HI or CO) shells, and discuss the various scenarios,
taking into account the 
predictions from N-body numerical simulations.

The observations are described in Sec. 2, the sample of shell galaxies in Sec. 3,
and the results are presented in Sec. 4. Finally, the interpretation and discussion
are gathered in Sec. 5, and the conclusions in Sec. 6.

\section{Observations and data reduction}
Observations of the $^{12}$CO(1-0) and $^{12}$CO(2-1) emission have been carried out at IRAM 30-m 
telescope at Pico Veleta, Granada, Spain, during 2017 July
 (for Arp~230, NGC~474 and NGC~3934), then September, October and 2000 November
  (for the remaining sources).
The beam full width at half-maximum (FWHM) is 23 arcsec and 11 arcsec at the frequencies 
of 115 GHz and 230 GHz, respectively. The SIS receivers (EMIR) were used for observations in the wobbler switching 
mode, with reference positions offset by $\pm$120 arcsec in azimuth. The main-beam efficiency of IRAM
is $\eta_{mb}=T_{A}^{*} / T_{mb}$ = 0.82 and 0.64 at 115 GHz and 230 GHz respectively. 
The system temperatures ranged
between 190 K and 420 K at 2.6 mm and between 240 K and 600 K at 1.3 mm.
The pointing was regularly checked every 2 h on a nearby planet or a bright continuum source, and the focus was 
reviewed after each sunrise or if a suitable planet was available, as well as at the beginning of each night.
The time on source typically ranged from 30 min to 1h, being weather-dependent. 
Two backends were used simultaneously, the autocorrelator WILMA, and the Fourier Transform
Spectrometer FTS. The rms noise level was
$\sigma _{mb}\sim$1.5mK [$T_{A}^{*}$] with a spectrometer resolution of 40 $km s^{-1}$ for 114 GHz and 
$\sigma _{mb}\sim$3.0mK [$T_{A}^{*}$] with a spectrometer resolution of 20 $km s^{-1}$ for 230 GHz.
The various pointings observed on the shell galaxies are indicated in Fig. \ref{Arp230-beams} to
\ref{a295n-beams}.

\begin{table*}
    \centering
\caption{Properties of the 9 shell galaxies}
	\begin{tabular}{lccccccccc}
\hline \hline
Name & RA & DEC & cz &  D  & M(HI) & L(IR) & M(H$_2$) & Ref & Type\\ 
  & J2000 & J2000 &\kms & Mpc  & logM$_\odot$ & logL$_\odot$   &  logM$_\odot$  &   & NED\\ 
\hline
Arp~230 &00:46:24.2&-13:26:32&1720&24.2&9.20&9.49&8.95&(1) & S0pec\\
NGC~474 &01:20:06.7&+03:24:55&2315&32.2&9.63&8.03&$<$7.7&(2) &SA0\\
Arp~10  &02:18:26.3&+05:39:14&9108&120.&10.2&10.5& -  &(3) & S?\\
NGC~3032&09:52:08.1&+29:14:10&1562&21.2&8.15&9.34&8.42&(4)& SAB0 \\
NGC~3656&11:23:38.6&+53:50:32&2870&42.6&9.27&9.99&9.67&(5) & I0pec\\
NGC~3934&11:52:12.5&+16:51:05&3779&49.2&9.51&9.98&9.32&(6) &S?\\
NGC~5018&13:13:01.0&-19:31:05&2816&37.5&8.81&9.54&$<$7.4&(7) & E3\\
NGC~7600&23:18:53.8&-07:34:50&3483&39.0&$<$7.5& -- & -  &(8) & S0\\
Arp~295N&23:42:00.8&-03:36:55&6966&93.0&10.2&10.9&10.2&(9) & Sbpec\\
\hline 
\end{tabular}
\label{tab:param}
\begin{list}{}{}
\item -- References (1) \cite{Schiminovich2013} for HI, \cite{Ueda2014} for CO --
	(2) \cite{Rampazzo2006} for HI, \cite{Combes2007} for CO --
	(3) \cite{Charmandaris1996} for HI  --
	(4) \cite{Lucero2013} for HI, \cite{Combes2007} for CO --
	(5) \cite{Balcells2001} for HI, \cite{Young2002} for CO --
	(6) \cite{Galletta1997} for HI and CO  --
	(7) \cite{Kim1988} for HI, \cite{Ueda2014} for CO --
	(8) \cite{Serra2008} for HI  --
  (9) \cite{Hibbard1996} for HI and CO  --
\item -- All H$_2$ data are centered and are representative of the total molecular content.
 L(IR) data are from NED.
\end{list}
\end{table*}                      

\section{The sample}

We have selected shell galaxies observable from the northern hemisphere,
which were previously observed in HI-21cm and in CO in the millimeter domain
(see Table \ref{tab:param}). In general they are quite rich in neutral gas,
except NGC~7600.
The properties of the 9 shell galaxies selected in our sample are gathered
in Table \ref{tab:param}. The H$_2$ masses in this Table are those which were known
before the present study, and all luminosities and masses have been scaled
to the adopted distances.
 References in this Table are for HI and H$_2$.
We give below the detailed properties of each system.

{\bf Arp 230}

\begin{figure}[h!]
\centerline{
\includegraphics[angle=0,width=7.5cm]{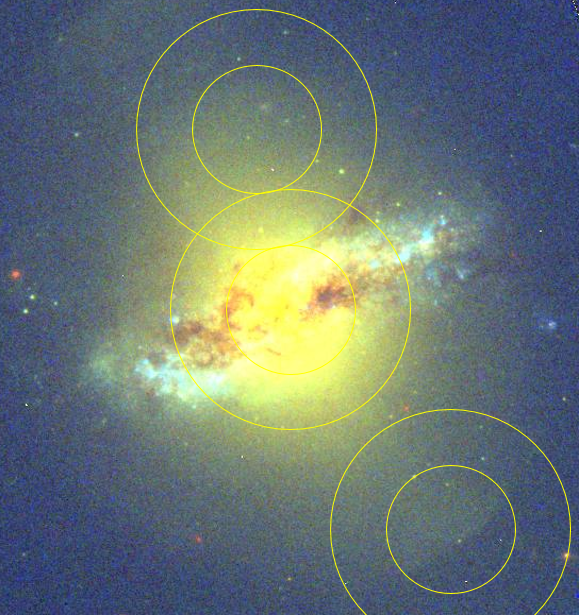}
}
\caption{The CO(1-0) 23'' and CO(2-1) 11'' beams observed
  towards Arp~230 are overlaid on an HST composite image (F336W, F555W, F814W).
   The two offset positions (NE and SW) are centered on shells.
}
\label{Arp230-beams}
\end{figure}

\cite{Arp1966} was the first to classify the galaxy as peculiar,
because of its faint stellar arcs, not yet called shells.
Arp 230 (IC51) is now considered to be one of the 
prototypes for shell galaxies formed around a prolate
object, with shells phase wrapped from the
radial infall of a small satellite \citep{Quinn1984}. 
The geometry of the shells is characteristic (see Fig. \ref{Arp230-beams}),
they are aligned with the major axis, and 
interleaved in their distance to the center \citep{Dupraz1986}.
At least eight shells can be counted,
with the most distant at 11~kpc, which is rather
rare, although not as complex as NGC~3923 \citep{Prieur1988}. 
As frequent in shell systems, the first shell of Arp 230 
(i.e. the most distant) is 
connected to the primary galaxy by a faint distorted tail or plume.

An HI mass of 1.59 10$^9$ M$_\odot$ has been detected in this merged system,
with the Green Bank 140foot \citep{Richter1994}.
With the VLA, \cite{Schiminovich2013} mapped the HI in Arp 230, 
with $\sim$0.8~kpc resolution.
 The HI is not aligned with the stellar shells, but perpendicular
 to the major axis, like a polar ring, aligned
 with the dust lane. The total HI mass detected with the VLA
 is 1.1 10$^9$ M$_\odot$.
\cite{McGaugh1990} together with \cite{Prieur1988} found that the color
of the shells are bluer than the main body of the Arp~230 galaxy,
suggesting that the merger provided gas and star formation to the shells.

\cite{Galletta1997} have detected CO(1-0) in the center with the Kitt Peak 12m telescope
(beam of 1 arcmin), and derived M(H$_2$) = 8.9 10$^8$ M$_\odot$.
The central dust lane corresponds to a molecular disk in rotation
in the center, mapped in CO(1-0) with ALMA at $\sim$ 0.5~kpc resolution \citep{Ueda2014}.
The corresponding H$_2$ mass is 1.7 10$^8$ M$_\odot$.

\smallskip
\noindent {\bf NGC 474/470 (Arp 227)}

\begin{figure}[h!]
\centerline{
\includegraphics[angle=0,width=7.5cm]{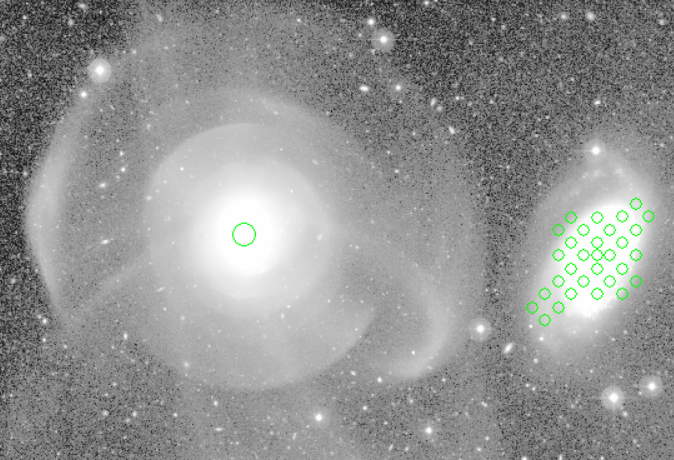}
}
\caption{Image in the g-band from the CFHT Megacam of NGC~474 and companion
NGC~470, enhancing the low surface brightness features, like the shells
	\citep{Duc2015}. The CO(1-0) 23'' beam is indicated at
	the center of NGC~474, while the maps of CO(2-1) 11'' beams 
	cover its companion (see Appendix).
}
\label{fig:n474-beams}
\end{figure}

The group has been observed in the optical by
\cite{Rampazzo2006, Rampazzo2007}, and with high sensitivity
by \cite{Duc2015}. In the latter CFH-Megacam image, NGC~474 appears
as a spectacular shell system, where shells are distributed
randomly at all azimuth (see Figure \ref{fig:n474-beams}). This geometry corresponds to stellar shells aligned 
in the equatorial band of an oblate galaxy \citep{Dupraz1986}. Superposed
to the shell systems, there is a radial tail/plume, coming from the past merger.

\begin{figure}[h!]
\centerline{
\includegraphics[angle=0,width=7.5cm]{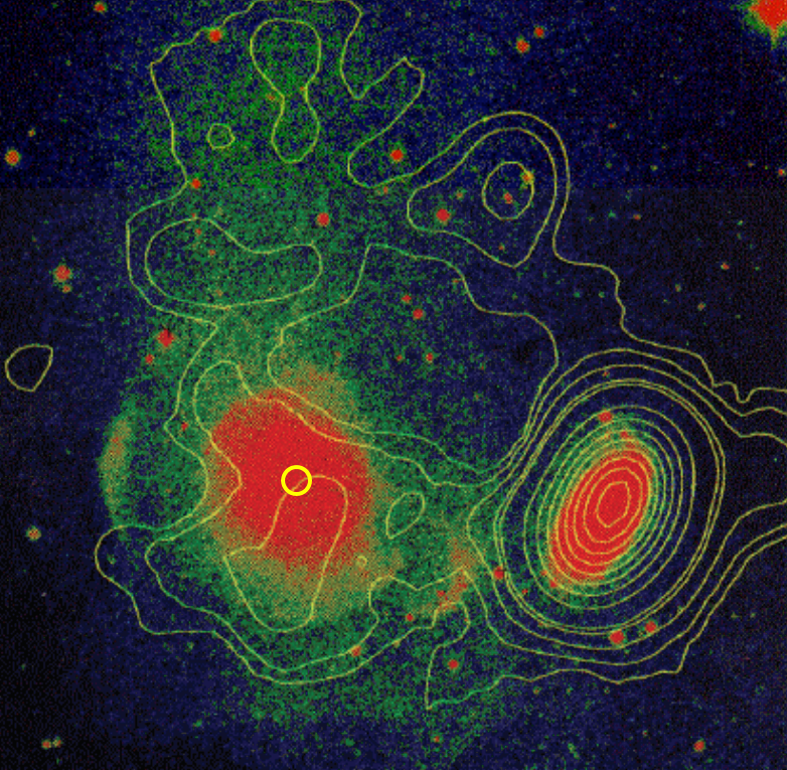}
}
	\caption{VLA HI map over Arp~227 (NGC~470 at right and NGC~474 at left),
  from \cite{Schiminovich1997} and van der Hulst priv. comm. The CO(1-0) 23'' beam is indicated
          with yellow circle.
}
\label{fig:HI-474}
\end{figure}

The first HI observations were not separating the two galaxies of the pair
\citep{Huchtmeier1982}. Interferometric observations with VLA have shown that most of the HI
reservoir belongs to the spiral NGC~470, but a long tail and bridge covers NGC~474,
and therefore the early-type shell galaxy may possess some interstellar gas 
\citep{Rampazzo2006}, see also Figure \ref{fig:HI-474}.

As early as 1993, \cite{Sofue1993} detected CO(1-0) emission towards 
Arp 227A (NGC 470), the late-type
of the pair, and derived an H$_2$ mass of 2.3 10$^8$ M$_\odot$.  
However, towards the shell early-type galaxy NGC~474, \cite{Combes2007} find an upper limit
of M(H$_2 <$ 3.8  10$^7$ M$_\odot$, with the IRAM-30m telescope.
 In the present work, we obtain a lower upper limit for the CO emission towards the center,
  and then did not point towards the shells, since the emission there is expected to be
  even smaller. We observed the companion NGC~ 470 to know how much
available gas mass existed there, that could have fallen towards NGC~474. 

\cite{Lim2017} studied the Globular Cluster (GC) population in the NGC~474
galaxy. GC and their color/age are good tracers of the merging history of
a system. The violent starbursts triggered by the major mergers of gas-rich
spirals favor the formation of metal-rich and red GC
\citep{Whitmore1993, Chien2007, Renaud2015}, while the dry minor 
merger of small-mass satellites brings relatively blue GC in the halo
of a massive primary. In the outer parts of NGC~474, there is a remarkable
spatial correlation between GC and stellar fine-structure (shells, tails, radial streams), while GC are blue, likely to have formed in low-metallicity
satellites, and accreted later. In the center of the galaxy, GC have
an intermediate color, compatible with their formation in situ, when 
star formation occurred 7-8 Gyr ago. With passive evolution until 13 Gyr, they 
will become red as expected in a massive elliptical today \citep{Lim2017}.

\smallskip
\noindent {\bf Arp 10}

\begin{figure}[h!]
\centerline{
\includegraphics[angle=0,width=7.5cm]{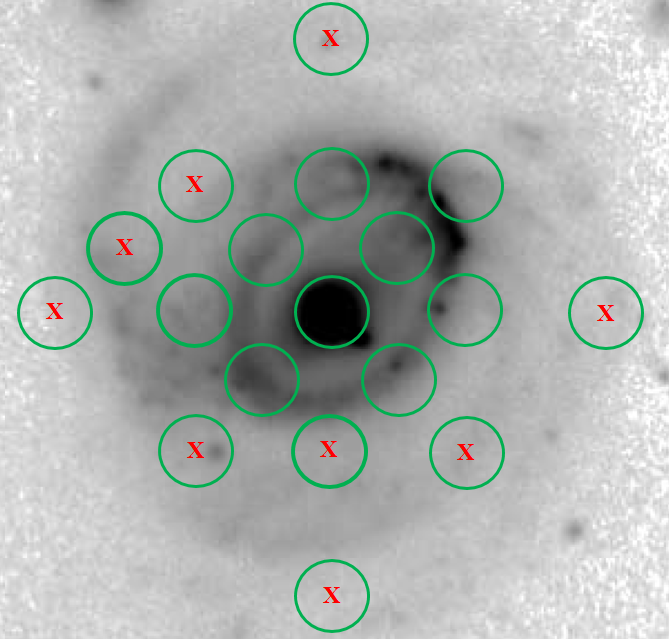}
}
\caption{CO(2-1)- 11 arcsec observed beams in Arp~10,
	overlaid on the blue optical image, from \cite{Bizyaev2007}.
	The 9 pointings selected to be outside the ring are indicated
	as a red cross (see Fig. \ref{arp10-stac}).
         They can all be considered to cover part of a shell.
}
\label{arp10-beams}
\end{figure}

This peculiar galaxy reveals loops, ripples and shells in its outer
parts \citep{Charmandaris1996}. The HI VLA map from the latter work
shows a neutral gas extension beyond the deepest optical shells.
The HI appears in two components: one more relaxed in a rotating disk
in the inner parts, although with non-circular motions, and the other 
component quite similar to a shell system, with no one-to-one 
correspondence with the optical shells.
\cite{Bizyaev2007} show that the inner parts of the optical disk are dominated
by a collisional ring, and the intruder responsible for the expanding
wave is found 5'' at the southwest from the nucleus of Arp10
(see Fig. \ref{arp10-beams}).
Its mass could be 1/4th of the Arp10's mass. Star formation has
been triggered in the ring by the collision.

\smallskip
\noindent {\bf NGC 3032}

\begin{figure}[h!]
\centerline{
\includegraphics[angle=0,width=7.5cm]{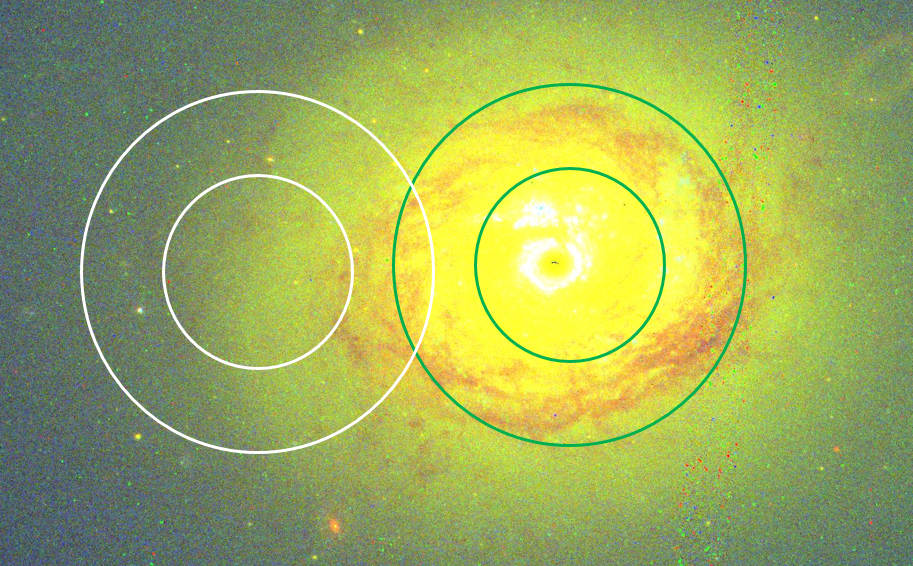}
}
	\caption{CO(1-0) and CO(2-1) observed beams on NGC~3032,
	  overlaid on an HST composite image).
           The eastern offset is centered on a shell.
}
\label{n3032-beams}
\end{figure}

This early-type galaxy is classified as a weakly barred lenticular.
Its optical image is very smooth, with however some dusty
inner disk, and a large-scale shell system, embedding the galaxy
(see Fig. \ref{n3032-beams} and \cite{Duc2015}).
 This galaxy has certainly recently acquired its gas, since the HI
 map shows counter-rotation with respect to the stars
 \citep{Oosterloo2010}. While all gas components (ionized, atomic and molecular)
 corotate, the main stellar body counter-rotate, but there is however 
 some young stars, probably formed since the gas accretion event, co-rotating with 
 the gas in the very center \citep{McDermid2006}.

\cite{Sage1989} detected CO(1-0) emission towards the peculiar 
lenticular NGC~3032 with the Kitt-Peak-12m telescope (60'' beam),
and derived an H$_2$ mass of 3.6 10$^8$ M$_\odot$,
computed with the same adopted distance and CO-to-H$_2$ conversion
factor than here. The latter is somewhat higher
than that found by \cite{Combes2007}, of 2.5 10$^8$ M$_\odot$,
but this could be explained, since the IRAM-30m beam is 23'' (2.4kpc)
in CO(1-0), and may miss some outer emission.
The CARMA interferometer mapped the inner disk, but not
covering more than the IRAM-30m beam \citep{Alatalo2013}.

\smallskip
\noindent {\bf NGC 3656}

\begin{figure}[h!]
\centerline{
\includegraphics[angle=0,width=7.5cm]{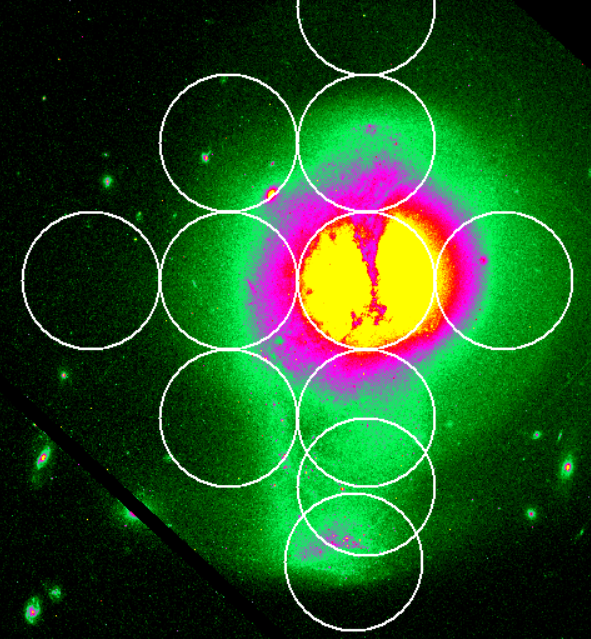}
}
	\caption{CO(1-0) 23'' beams observed towards NGC~3656, superposed
	on an F606W HST image. Note the prominent N-S dust lane, and the southern 
	shell at the bottom of the image.
 All CO beams are partially covering some shell, except the central one,
	which includes the dust lane.
        }
\label{n3656-beams}
\end{figure}

This peculiar galaxy (Arp 155) has been classified
as a shell elliptical galaxy, with several shells, and one particularly
strong 9~kpc south of the center \citep{Balcells2001},
as can be seen on Fig. \ref{n3656-beams}.
Two kinematical components have been identified, with perpendicular axes
\citep{Balcells1990}. The inner (1kpc) stellar core rotates with a major axis E-W, and 
the larger stellar and gaseous disk (radius 7 kpc) rotates with a major axis N-S.
The VLA HI map from \cite{Balcells2001} shows two gas components:
an edge-on and warped inner disk extending up to 7kpc (M(HI)=2 10$^9$ M$_\odot$), 
aligned north-south along an inner dust lane; and a more extended outer
structure, distributed in shells or rings, roughly correlated
to the optical shells, but more extended. The HI southern shell at 9~kpc
coincides with the stellar shell, and has the systemic velocity.

\cite{Wiklind1995} detected CO emission towards the merger system NGC~3656
and derived an H$_2$ mass of 2.2 10$^9$ M$_\odot$. A CO map was done with the
BIMA interferometer by \cite{Young2002} and revealed the same N-S alignment
as the HI disk. The map was not sensitive enough to detect CO in the southern shell.

\smallskip
\noindent {\bf NGC 3934}

\begin{figure}[h!]
\centerline{
\includegraphics[angle=0,width=7.5cm]{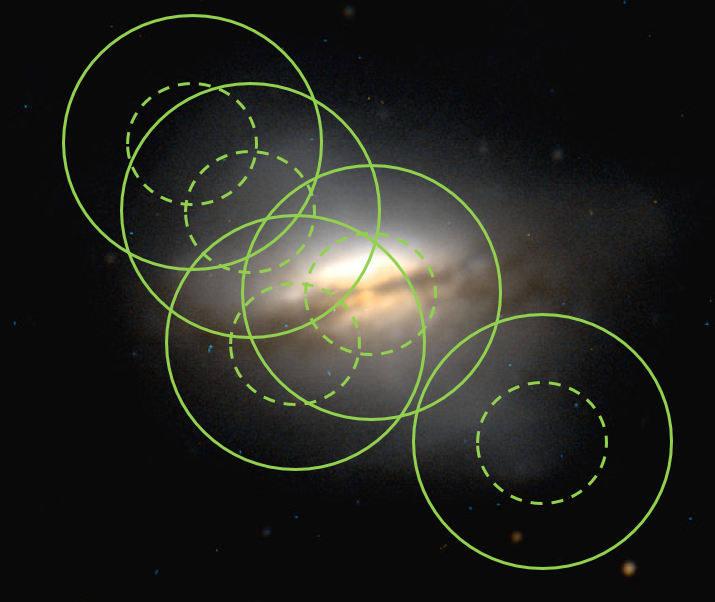}
}
	\caption{CO(1-0) 23'' and CO(2-1) 11'' beams observed towards NGC~3934,
	overlaid on a composite optical image from the Sloan Digital Sky Survey (SDSS).
All offset beams are covering some shells, which are expected to show plateau emission before
  a sharp edge.
}
\label{n3934-beams}
\end{figure}

This peculiar disk galaxy is a candidate polar ring \citep{Whitmore1990},
belonging to a compact group, similar to 
an Hickson group \citep{Bettoni2011}. It has a strong linear dust 
lane in the center,
that irregularly obscures the inner and brighter regions of the galaxy
(see Fig. \ref{n3934-beams}). Spurs and shells are visible in the inner parts, but 
especially in the outskirts of the galaxy. The latter have an average B - R $\sim 1.2$,
not as blue as typical spiral
arms although they are bluer than the central parts reddened by the dust-lane. 
The galaxy has then a rich shell system, dispatched all around the center,
in apparent random order. This shell morphology can be reproduced in models
with oblate galaxies, with shells near the equatorial plane \citep{Dupraz1986}. 
NGC~3934 is rich in gas, it was detected in HI, with a mass of 5 10$^9$ M$_\odot$
\citep{vanDriel2000} and in CO \citep{Galletta1997} 
with M(H$_2$) = 2.1 10$^9$ M$_\odot$. It  has a star formation
rate of SFR = 1.6 M$_\odot$/yr, derived from its far-infrared IRAS flux.
\cite{Bettoni2011} propose a formation in a dissipative merger, through
hydrodynamical SPH simulations. From this formation mechanism, molecular
gas might be expected in the shells.

\smallskip
\noindent{\bf NGC 5018}

Like NGC~7252, NGC~5018 is one of the prototypes of a recent merger, 
3Gyr ago, which is
becoming a giant elliptical (gE) \citep{Buson2004}. It is the dominant
gE of a small group. It has been detected by IRAS, with an SFR = 0.7 M$_\odot$/yr.
The galaxy reveals sharp outer shells \citep{Rampazzo2007}, and a central
linear dust lane.
\cite{Kim1988} mapped the HI, and found M(HI) = 6.1 10$^8$ M$_\odot$. 
On the contrary, the galaxy is not detected in CO(1-0) with ALMA
\citep{Ueda2014}, with M(H$_2$) $<$ 2.3 10$^7$ M$_\odot$, but is detected
in 3mm continuum.

\smallskip
\noindent {\bf NGC 7600}

This is a typical shell galaxy, with a large number of shells,
aligned and interleaved along the major axis. The main body is
a flattened early-type galaxy, classified as a 
lenticular. \cite{Turnbull1999} count 16 shells, with the first one at 40 kpc from
the center. \cite{Cooper2011} reveal some even deeper external features,
and reproduce easily the system of shells from a simulation 
of phase wrapping during the radial merger of a small companion.
There is only an upper limit for HI in this galaxy \citep{Sansom2000, Serra2008}.
It was selected in our sample, however, because of its high degree 
of fine stellar structures.

\smallskip
\noindent {\bf Arp~295N}

\begin{figure}[h!]
\centerline{
\includegraphics[angle=0,width=7.5cm]{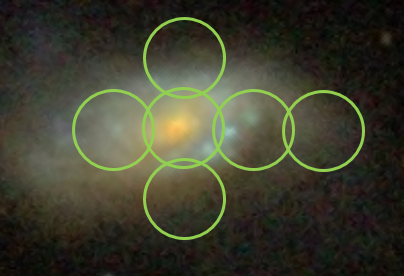}
}
\caption{CO(2-1) 11'' beams observed towards Arp~295N,
  overlaid on a composite SDSS image.
            The western offsets are covering shells.
}
\label{a295n-beams}
\end{figure}

The system Arp~295 is a pair of galaxies in an early stage of merger.
The Northern member reveals large plumes, shells and loops of H$\alpha$
emission up to 9 kpc north and south of the minor axis
(see Fig. \ref{a295n-beams}).
The northern galaxy is undergoing a retrograde encounter
with its southern companion. The HI was mapped with the VLA by
\cite{Hibbard1996}, who found an HI mass of 1.7 10$^{10}$ M$_\odot$ and
an SFR = 17 M$_\odot$/yr. They also report about our IRAM-30m observations
of CO emission, which were not published, with a molecular mass of
M(H$_2$) = 8.5 10$^9$ M$_\odot$. The star formation rate of the ensemble
has been estimated to be 21.9  M$_\odot$/yr from Pa$\alpha$ 
by \cite{Tateuchi2015}.

\section{Results}

In each shell galaxy, the center was observed first, to check the presence
of molecular gas, and if present, the shell regions were also observed. Indeed,
the presence of gas in shells is expected to come from the infall of a gas-rich
satellite \citep{Charmandaris2000, Combes2000}. The gas present in the satellite
is first liberated and may oscillate in the potential of the primary galaxy, just before
the satellite stars form the shells. Some gas is expected to fall in the center through
dissipation. The presence of gas in the center is therefore a necessary condition
to find gas in shells, but is not sufficient, since the primary galaxy could also have had
some gas before the collision. Out of the 9 shell galaxies of our sample,
3 were not detected in their center (NGC~474, NGC~5018 and NGC~7600)
and they were not searched for gas in their shells. The upper limits are
displayed in Table \ref{tab:CO-results}. For the 6 remaining shell galaxies,
3 were mapped extensively, and we show spectra maps in CO(1-0) and CO(2-1) for Arp~10, NGC~3656
and Arp~295N, in Fig. \ref{fig:mapa10}, \ref{fig:mapa10_2}, \ref{fig:map36}, 
\ref{fig:mapa295n}, \ref{fig:mapa295n_2}, respectively.  In Table \ref{tab:CO-results}
only the central beam, and the offset beams covering a shell and detected
in CO(1-0), are characterized.
For the 3 others (Arp~230, NGC~3032 and NGC~3934), only a few spectra are shown, in
Fig. \ref{fig-a230}, \ref{fig-3032} and \ref{fig-3934}. All these points are detailed in
Table \ref{tab:CO-results}.

\subsection{Molecular gas content}
\label{sec:content}

To quantify the amount of molecular gas found in each pointing, we rely on
the empirically established proportionality between the CO(1-0) luminosity and
H$_2$ mass, for a large number of Milky-Way like galaxies \citep[e.g.,][]{Bolatto2013}.
We convert the integrated intensities in each beam (in T$_{mb}$) in integrated fluxes S(CO) dV, with the
conversion of 5 Jy per Kelvin, appropriate for the IRAM-30m telescope. 
From the integrated flux S(CO) dV (Jy km/s) found within each region, the derived molecular mass
is obtained through the formula
$$
{\rm L'}_{\rm CO} (K km/s/pc^2) = 3.25 x 10^7 \frac{S(CO) dV}{(1+z)}  \left(\frac{D_L}{\nu_{rest}}\right)^2, 
$$
where $\nu_{rest}$ =115.271 GHz, and D$_L$ is the luminosity distance in Mpc
\citep[e.g.,][]{Solomon2005}. The molecular mass, including helium, is then
derived from
$$
{\rm M}(H_2) = 4.36 {\rm L'}_{\rm CO} 
$$
assuming the standard CO-to-H$_2$ conversion factor of X$_{CO}$ = 2~10$^{20}$ cm$^{-2}$(K~km/s)$^{-1}$,
 applicable to Milky Way-like galaxies.
 Although shell galaxies are interacting, they are mildly perturbed, and certainly not starbursts,
 or Ultra-Luminous InfraRed Galaxies (ULIRG), where a lower conversion factor might apply.

 Molecular masses, corresponding to each CO(1-0) pointing are displayed in
 Table \ref{tab:CO-results}. In most cases, we have not observed the whole galaxy disks,
 and an estimation of the total molecular masses would need some extrapolation.

\begin{table*}[htb]
\caption{CO results}
\centering
\label{tab:CO-results}
\begin{tabular}{rccccccccc}
\noalign{\smallskip} \hline \hline \noalign{\smallskip}
	&  &  &  CO(1-0) &  & &  CO(2-1) & & &\\
Galaxy & Offsets & Int-Flux & V &$\Delta$V & Int-Flux  & V  & $\Delta$V & M(H$_2$) &$\Sigma$(H$_2$) \\
& ('','')   & Jy\kms & \kms & \kms & Jy\kms & \kms & \kms  & $10^8$ M$_\odot$&M$_\odot$pc$^{-2}$\\
 \hline
ARP~230&(0,0)&9.2$\pm$0.9 &-54$\pm$6&125$\pm$15& 14.7$\pm$1&-42$\pm$5&111$\pm$11&0.57$\pm$0.06&11\\
& (4.4,17)& 3.7$\pm$0.7 & 5$\pm$17& 190$\pm$40& $<$3 & -- &  --  & 0.23$\pm$0.04&4\\
  & (-14,-21)& $<$1.6  & -- & -- & $<$1.7 & -- &  --  & $<$0.1& $<$2\\
\hline
NGC~474  & (0,0)& $<$1.4  & -- & -- & $<$3.1 & -- &  --  & $<$0.15&$<$1.6\\
 \hline
ARP~10&(0,0)&29.7$\pm$1 &10$\pm$4&208$\pm$7& 38.8$\pm$2.7&25$\pm$8&224$\pm$16&44.6$\pm$1.5&35\\
      & see &18 pt& map&  & & & & &\\ 
\hline
NGC~3032&(0,0)&71$\pm$1.3 &57$\pm$1&129$\pm$2&86.5$\pm$2&47$\pm$1&108$\pm$3&3.6$\pm$0.06&90\\
& (20,0)& 8$\pm$0.8 & 80$\pm$3& 53$\pm$6& $<$5 & -- &  --  & 0.4$\pm$0.04&10\\
 \hline
NGC~3656&(0,0)&119.6$\pm$2.2&-34$\pm$4&433$\pm$8 &200$\pm$4.5&-31$\pm$7&407$\pm$23&22.9$\pm$0.41&142\\
& (22,22)& 9.2$\pm$1.3&-192$\pm$8&112$\pm$17 &29.8$\pm$4.0&6$\pm$10&322$\pm$45&1.7$\pm$0.25&11\\
& (0,22)& 9.4$\pm$1.6&-184$\pm$12&153$\pm$33 &$<$8  &  --  &  --  &1.8$\pm$0.3&11\\
&(22,0)& 47.6$\pm$2.0&-20$\pm$9&433$\pm$18 &2.7$\pm$0.8&-78$\pm$8&51$\pm$19&9.1$\pm$0.4&57\\
& (22,-22)&8.6$\pm$1.7&107$\pm$14&143$\pm$32 &5.7$\pm$1.0&106$\pm$6&73$\pm$14&1.64$\pm$0.3&10\\
&(0,-22)&18.6$\pm$1.5&130$\pm$9&232$\pm$23 &$<$7  &  --  &  --  &3.55$\pm$0.3&22\\
&(0,-33)&3.6$\pm$1.0&126$\pm$18&135$\pm$46 &4.8$\pm$1.0&91$\pm$16&149$\pm$30&0.69$\pm$0.2&4.3\\
&(2,-45)&1.5$\pm$0.3&179$\pm$15&139$\pm$29 &2.5$\pm$0.8&105$\pm$34&187$\pm$67&0.29$\pm$0.06&1.8\\
& see &11 pt& map&  & & & & &\\ 
\hline
NGC~3934&(0,0)&42.8$\pm$0.8&22$\pm$4&384$\pm$7&78.9$\pm$1.7&30$\pm$4&396$\pm$9&11$\pm$0.2&51\\
    &(7,-5)&21.4$\pm$0.8&-17$\pm$7&350$\pm$13&18.7$\pm$3&-78$\pm$19&210$\pm$67&5.5$\pm$0.2&25\\
    &(16,14)&3.7$\pm$0.4&-128$\pm$7&115$\pm$17&1.2$\pm$0.4&-119$\pm$17&86$\pm$27&0.9$\pm$0.1&4\\
  &(-16,-14)&6.6$\pm$1&134$\pm$18&255$\pm$77&3$\pm$0.5&184$\pm$7&90$\pm$19&1.7$\pm$0.2&8\\
  & (11,12)& $<$10  & -- & -- & $<$90 & -- &  --  & $<$2.5&$<$11\\
\hline		 
 NGC~5018  & (0,0) & $<$1.8  & -- & -- & $<$4 & -- &  --  & $<$0.27&$<$2\\
   & (44,-22) & $<$3  & -- & -- & $<$4 & -- &  --  & $<$0.45&$<$3.6\\
   & (-22,22) & $<$3  & -- & -- & $<$3 & -- &  --  & $<$0.45&$<$3.6\\
\hline		 
 NGC~7600  & (0,0) & $<$3  & -- & -- & $<$4.8 & -- &  --  & $<$0.5&$<$3.6\\
   & (143,32) & $<$2  & -- & -- & $<$5 & -- &  --  & $<$0.3&$<$2\\
 \hline
ARP~295N&(0,0)&58$\pm$4&-54$\pm$11&333$\pm$24& 38$\pm$5&-42$\pm$20&268$\pm$40&53$\pm$4&69\\
&(-10,0)&45$\pm$3.8&16$\pm$14&340$\pm$33& 25.8$\pm$5.6&85$\pm$14&136$\pm$35&41$\pm$3.4&53\\
&(-20,0)&7.6$\pm$2.2&110$\pm$13&93$\pm$35& $<$2 & -- &  -- &6.9$\pm$2&9\\
& see &6 pt& map&  & & & & &\\ 
\noalign{\smallskip} \hline \noalign{\smallskip}
\end{tabular}
\begin{list}{}{}
\item --Integrated flux, velocity and FWHM obtained through gaussian fits 
      -- Velocities are counted from the central redshifts indicated
	in Table \ref{tab:param}
      -- The H$_2$ masses are estimated within the CO(1-0) beam of 23'', with
	the standard conversion factor of the Milky Way $\alpha$ = 
	4.36~M$_\odot~({\rm K}~{\rm km}~{\rm s}^{-1}~{\rm pc}^{2})^{-1}$.
-- Upper limits are given at 3$\sigma$, assuming a FWHM of 300\kms.
\end{list}
\end{table*}

The last column of Table \ref{tab:CO-results} reports the average molecular surface
density, within the CO(1-0) beam of 23 arcsec for each galaxy. This shows that galaxies
of later type (like NGC~3656 or NGC~ 3934) tend to have a richer content in molecular gas,
which might be due to their own gas before the accretion event.
 We have checked that the total molecular content is correlated
 to the FIR luminosity (see Table \ref{tab:param}), and that galaxies of our sample 
 are compatible with the global Kennicutt-Schmidt (KS) diagram, as shown
 for local early-type galaxies by \cite{Combes2007}. However,
 we have not enough sensitivity in H$\alpha$ or FUV/NUV (GALEX) images to be 
 able to determine the star formation surface density on shells, and
 compare to the molecular surface densities, to study the KS diagram of shells.

\subsection{CO Spectra}

For most of the shell galaxies detected in CO emission at least in
their center, we have tried to cover the outer stellar shells,
through simultaneous pointing in CO(1-0) and CO(2-1) lines.
 Shells are not only reduced to their sharp edges, but extend also inside
  these arcs in a nearly flat plateau emission, as can be seen in Figure \ref{fig:n474-beams}
  for example. With the large CO beams, we can thus hope to collect also emission from
these regions.
The best resolution we have, 11 arcsec, corresponds to 1-5 kpc typically on the galaxies, and is
just sufficient to separate the shell regions from the center, while the 23 arcsec beam is not.
Frequently, the CO(1-0) emission is detected in the shell offset position, but the CO(2-1) is not,
which can be interpreted as the absence of detection in the shell itself. The CO(1-0) emission
is then only due to some central emission seen at the border of the beam.

\begin{figure}[h!]
\centerline{
\includegraphics[angle=0,width=8cm]{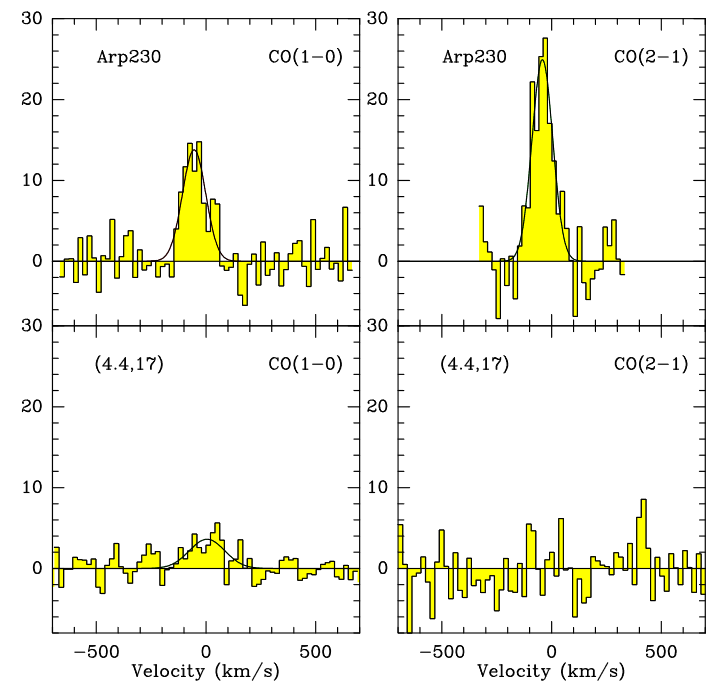}
}
\caption{CO spectra in Arp 230, central position and offset 
(in arcsec). The vertical axis is main beam temperature in mK. }
\label{fig-a230}
\end{figure}

This is clearly the case for Arp~230, as shown in Fig. \ref{fig-a230}. There is a strong
concentration of CO emission towards the center, which results in the peak main beam
temperature being nearly twice in CO(2-1) with respect to CO(1-0). The offset beam, at
17.5 '' = 2~kpc from the center, is detected in CO(1-0) but is consistent with emission from the central disk.
The offset is not detected in CO(2-1), which beams does not overlap with the disk.
\cite{Ueda2014} have mapped the CO(1-0) emission with ALMA with 4 arcsec resolution, and
showed that it is confined in a thin rotating edge-on disk, 5kpc in diameter, corresponding to the dust lane
prominent in Fig. \ref{Arp230-beams}.
Since no emission at all is detected in the farthest shell, at 3kpc from the center, we conclude that
there is no gas detectable in the shells.

\begin{figure}[h!]
\centerline{
\includegraphics[angle=0,width=8cm]{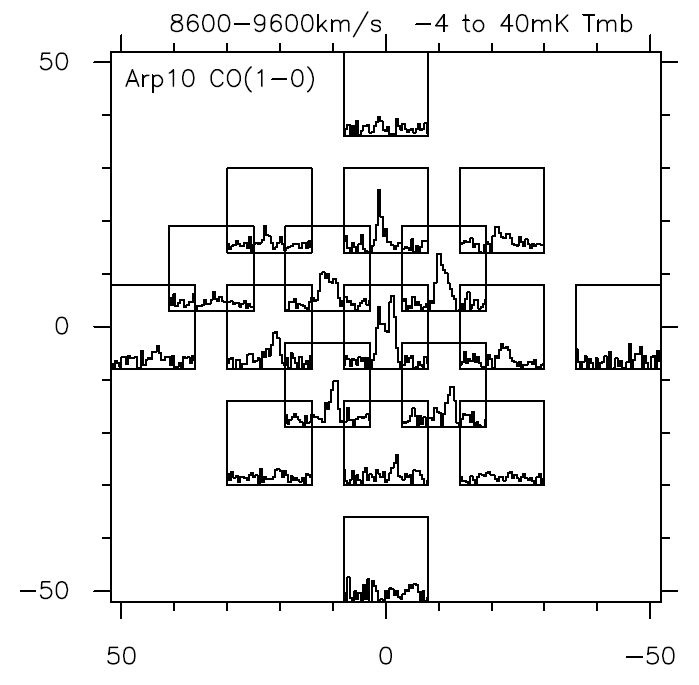}
}
\caption{CO(1-0) map of spectra in Arp~10. The RA and DEC
offsets are in arcsec. The scales of all spectral boxes,
in velocity and main beam temperature, are indicated on top.
}
\label{fig:mapa10}
\end{figure}

\begin{figure}[h!]
\centerline{
\includegraphics[angle=0,width=8cm]{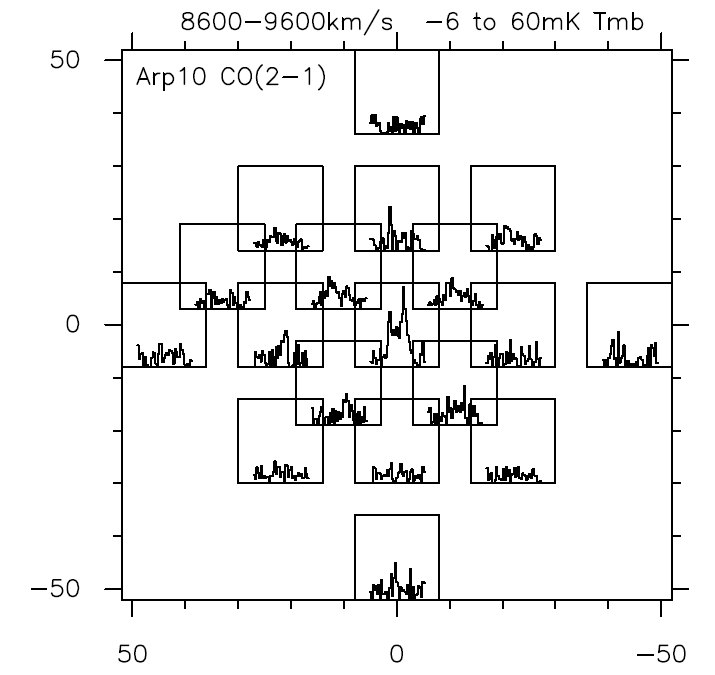}
}
\caption{CO(2-1) map of spectra in Arp~10.
}
\label{fig:mapa10_2}
\end{figure}

\begin{figure}[h!]
\centerline{
\includegraphics[angle=0,width=7cm]{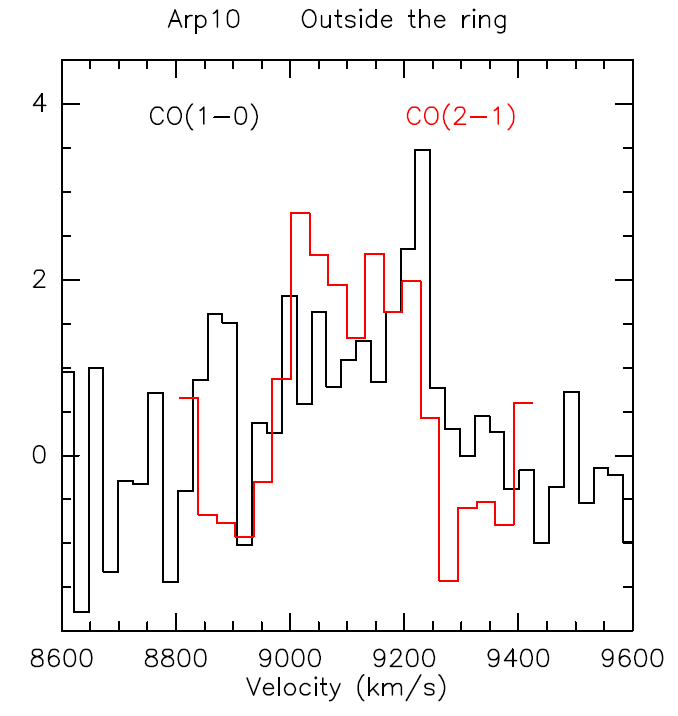}
}
\caption{Stacked CO spectra in Arp~10, considering only the positions
	outside the ring,  covering the shells (see Fig. \ref{arp10-beams}). 
The vertical axis is main beam temperature in mK. }
\label{arp10-stac}
\end{figure}

\begin{figure}[h!]
\centerline{
\includegraphics[angle=0,width=8cm]{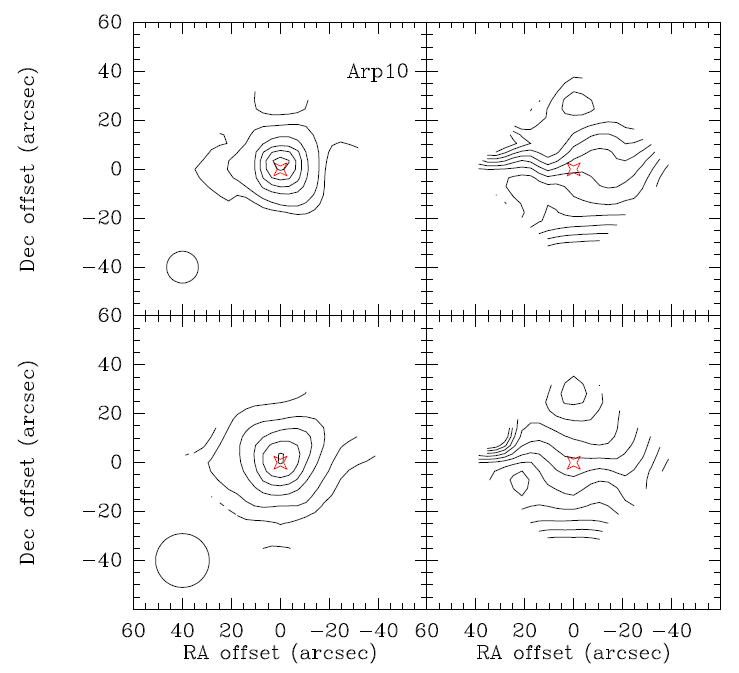}
}
\caption{ Isocontours of the integrated intensities in Arp~10 for CO(2-1)
  (Upper panels) and CO(1-0) (Lower panels), with linear contours spaced by
  1.52 and 1.15 K.\kms respectively, and maximum contour at 7.6 and 5.7 K.\kms
  respectively.
        The right panels display the corresponding velocity fields. The isovels
	are distant by 30 \kms, they start from 8990\kms in the North,
	to 9290\kms in the South. The beam sizes are indicated at the bottom left.
}
\label{fig:carta10}
\end{figure}

The case of Arp~10 is different, since the collision at the origin of the shells,
also produced a collisional ring in the center, i.e. a wave propagating radially outward,
from the center, due to the excitation of a head-on collision with a small impact parameter
\citep[e.g.,][]{Appleton1996}. In the CO(2-1) map of spectra of Fig. \ref{fig:mapa10_2},
the emission is clearly extended, not only on the ring, but also on the shells. The comparison
between the two lines (Fig. \ref{fig:mapa10}  and \ref{fig:mapa10_2}) suggests that indeed,
the gas has been detected as far as $\sim$ 26 kpc from the center, in ripples and shells
due to the interaction. To better quantify the amount of gas involved, we have stacked
all the spectra clearly outside the ring: the 9 offsets considered are indicated by a 
red cross in Fig. \ref{arp10-beams}.
 These 9 offsets correspond to the shells.
The resulting stacked spectra of CO(1-0) and CO(2-1) 
lines are plotted in Fig. \ref{arp10-stac}.  All these positions outside the ring were
  stacked, with their velocity recentered  around the systemic velocity of 9100 \kms.
  The resulting CO(2-1) spectrum
is the most significant, due to the smaller beams, excluding overlap with the center.
The strength of the CO(2-1) is similar and even stronger than the CO(1-0), meaning
the presence of molecular gas outside the ring, in the shells. From this stacked
spectrum, we derived a molecular content of M(H$_2$) = 3.4 10$^9$ M$_\odot$ in the shells, 
assuming the conversion factor adopted in Sec. \ref{sec:content}.
 An overview of the CO(1-0) and CO(2-1) surface densities and velocity fields
 is displayed in Fig. \ref{fig:carta10}. The velocity gradient is comparable to that
 of the ionized gas, and is limited
 due to the almost face-on (i=22$^\circ$) inclination \citep{Bizyaev2007}.

\begin{figure}[h!]
\centerline{
\includegraphics[angle=0,width=8cm]{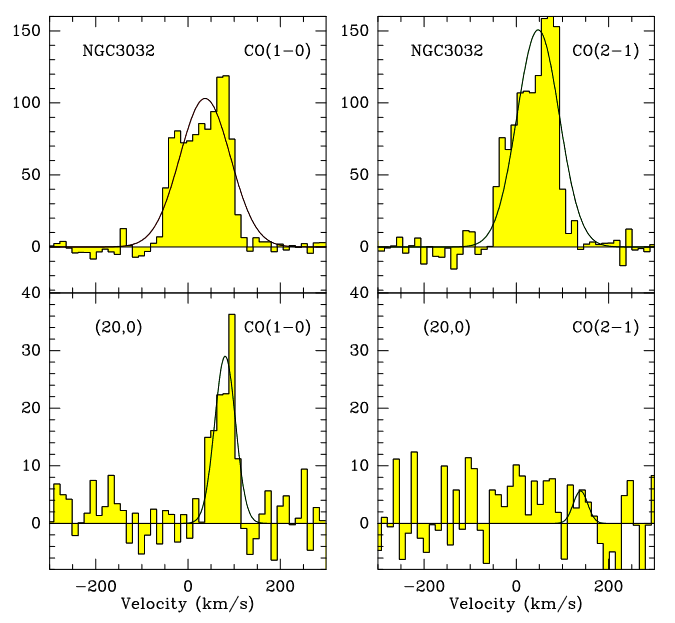}
}
\caption{CO spectra in NGC~3032, central position and offset 
(in arcsec). The vertical axis is main beam temperature in mK. }
\label{fig-3032}
\end{figure}

The case of NGC~3032 is quite similar to the map shown in Arp~230.  Fig. \ref{fig-3032} is typical of a quite
concentrated CO emission, with the main beam temperature of the CO(2-1) line
peaking higher than that of the CO(1-0).  The offset shell region is not overlapping the center
with the 11 arcsec beam, and no CO(2-1) is detected there, while the CO(1-0) line shows
the redshifted part of the emission, compatible with the inner rotating disk, expected
to coincide with the dust lanes (see Fig. \ref{n3032-beams}).  It can be concluded that the shell
region is not detected in the molecular gas.

\begin{figure}[h!]
\centerline{
\includegraphics[angle=0,width=8cm]{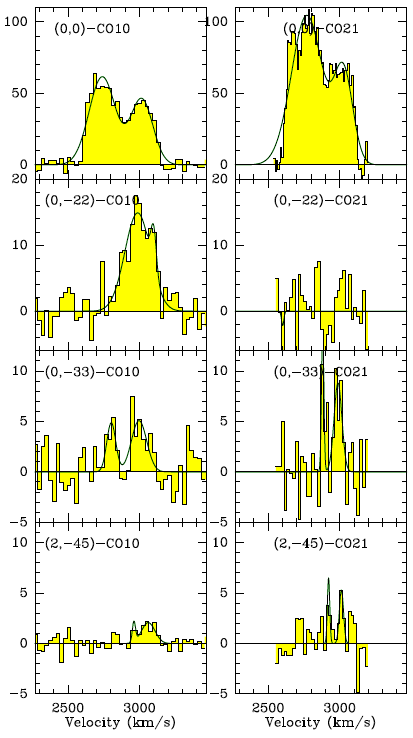}
}
\caption{CO spectra in NGC~3656, for the center and the offsets 
  (in arcsec) covering the southern shell.
  The vertical axis is main beam temperature in mK.}  
\label{fig-spec7b}
\end{figure}
\begin{figure}[h!]
\centerline{
\includegraphics[angle=0,width=8cm]{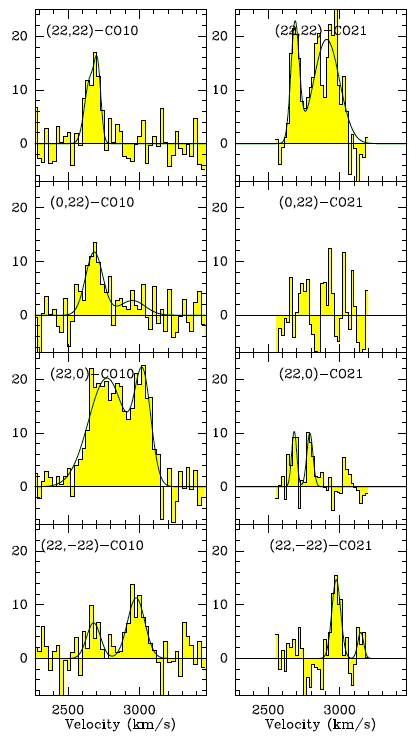}
}
\caption{Same as fig \ref{fig-spec7b} for only the offsets of NGC~3656
	covering a shell (other than the southern one), and detected in CO(1-0).}
\label{fig-spec7}
\end{figure}

\begin{figure}[h!]
\centerline{
\includegraphics[angle=0,width=8cm]{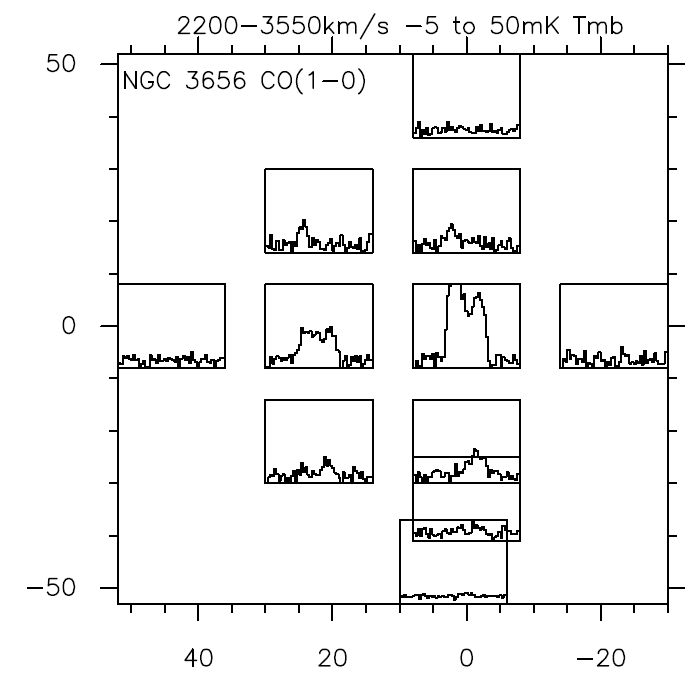}
}
\caption{CO(1-0) map of spectra in NGC~3656.
  The CO emission is most visible in the southern shell,
  but also in the three other directions.
}
\label{fig:map36}
\end{figure}

The shell galaxy NGC~3656 appears to be one of the best case of an
elliptical galaxy having accreted a companion, and having formed shells through
phase-wrapping \citep{Balcells2001}. The accreted companion must have been rich in gas,
having left a N-S dust lane, terminated by the southern shell at 9~kpc.
 This southern shell is clearly detected in molecular gas, 
 as seen through the comparison of the CO(2-1) and
CO(1-0) spectra in Fig. \ref{fig-spec7b}, and corresponding map
(Fig. \ref{fig:map36}). 
 There are also shells in the outer north, east and west observed beams, which show
 CO(1-0) emission, but no clear CO(2-1) (Fig. \ref{fig-spec7} and Table \ref{tab:CO-results}).
 CO(1-0) and CO(2-1) are simultaneously clearly detected only in the (22,22),
 (22,-22) and (0,-33) positions.
The velocity of the CO gas in the shells spans a range
around the systemic velocity, 
as is the velocity of stars in the southern shell \citep{Balcells2001}.
 In the extremity of the southern shell, the velocity profile is broad:
 it shows a double-horn shape from 2700 to 3100 \kms at (0,-33)
 and a single component from 2900 to 3100 \kms at (2,-45) in CO(1-0).
 In other places, it also shows a two-horn shape,
 which usually is only seen in the central beam for rotating disks, but is
 not compatible with a rotating disk in the outer parts. This means that there
 exist two components, with different angular momentum. In the HI component,
 there is also a trend of counter-rotation, as claimed by \cite{Balcells2001}.
The inner disk is however rotating, being redshifted in the south, and blueshifted in the north,
as is the HI disk, observed by \cite{Balcells2001}.
 The fact that the CO emission observed toward the southern shell has a velocity
  different from what is expected from the disk rotation, 
  confirms its origin in a phase-wrapped shell.

Averaging over the 3 observed beams covering the southern shell,
we obtain a surface density of $\Sigma$ = 9 M$_\odot$pc$^{-2}$ and we
derive a molecular content of M(H$_2$) = 3.8 10$^8$ M$_\odot$ in the 9~kpc-shell,
assuming the conversion factor adopted in Sec. \ref{sec:content}.

\begin{figure}[h!]
\centerline{
\includegraphics[angle=0,width=8cm]{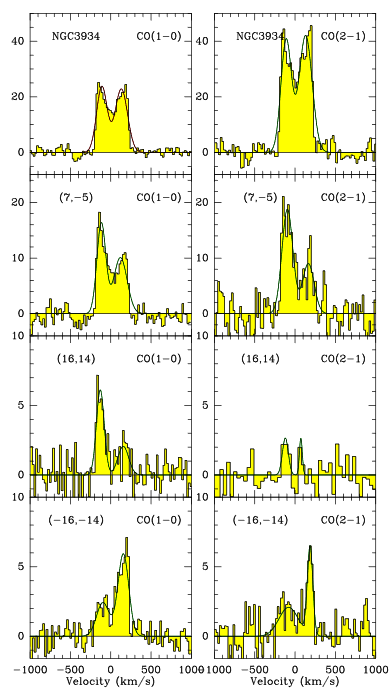}
}
\caption{CO(1-0) (left) and CO(2-1) (right) spectra in NGC~3934, central position and offsets 
  (in arcsec). The vertical axis is main beam temperature in mK.
   The two last offsets are covering the outer shells, without contamination from the
    center.
}
\label{fig-3934}
\end{figure}

NGC~3934 is a late-type galaxy, with randomly oriented shells around
a nearly edge-on disk, traced by a linear dust lane. It has been confused
with a polar ring galaxy, but this is not the right classification \citep{Bettoni2011}.
The beams clearly pointed to the shells, and not overlapping with the disk
(see Fig. \ref{n3934-beams}), with symmetrical offsets of 21'' = 5 kpc from
the center, are clearly detected in both CO(1-0) and CO(2-1),
as seen in Fig. \ref{fig-3934}. Therefore molecular gas is detected in the shells.
There is no shift in average velocity for the gas in the shells, with respect to the disk.
 However, the line profiles in the shells show a two-horn shape, indicating the 
presence of two opposite signs of the relative angular momentum of the shell material
relative to the primary galaxy.  This two-horn shape is frequently observed towards
the center of galaxies, and are then characteristic of a disk in rotation. This is due
to the observing beam covering both the receding (red) and approaching (blue) sides.
In the outer parts, only one velocity component is expected for a rotating disk.
The presence of a two-horn shape in the outer profiles supports the gas shell scenario.
  The varying CO(2-1) to CO(1-0) ratios from one position to the other is due
  to the different beams, and help to distinguish the possible contamination in
  the larger CO(1-0) beam.

Summing the molecular content of the two shells detected, we find
 M(H$_2$) = 2.6 10$^8$ M$_\odot$, see Table \ref{tab:CO-results}.

\begin{figure}[h!]
\centerline{
\includegraphics[angle=0,width=8cm]{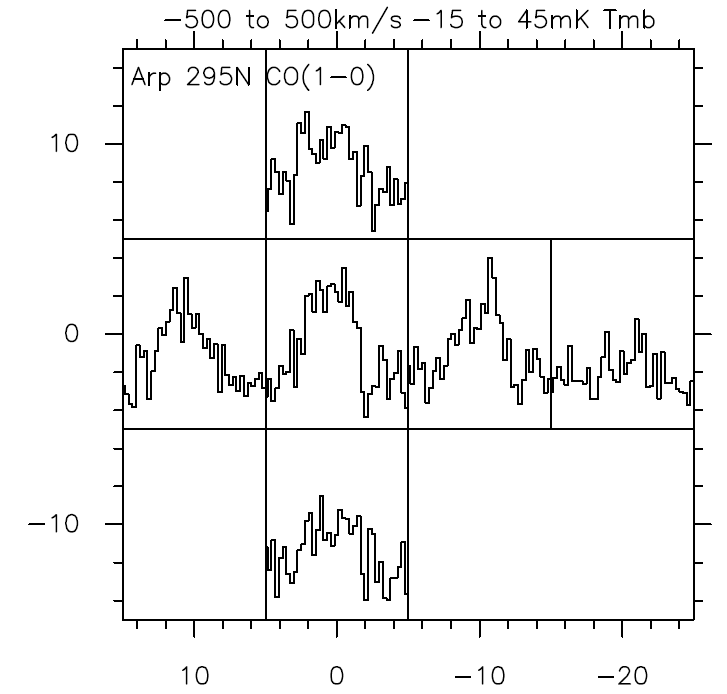}
}
\caption{CO(1-0) map of spectra in Arp~295N.
}
\label{fig:mapa295n}
\end{figure}

\begin{figure}[h!]
\centerline{
\includegraphics[angle=0,width=8cm]{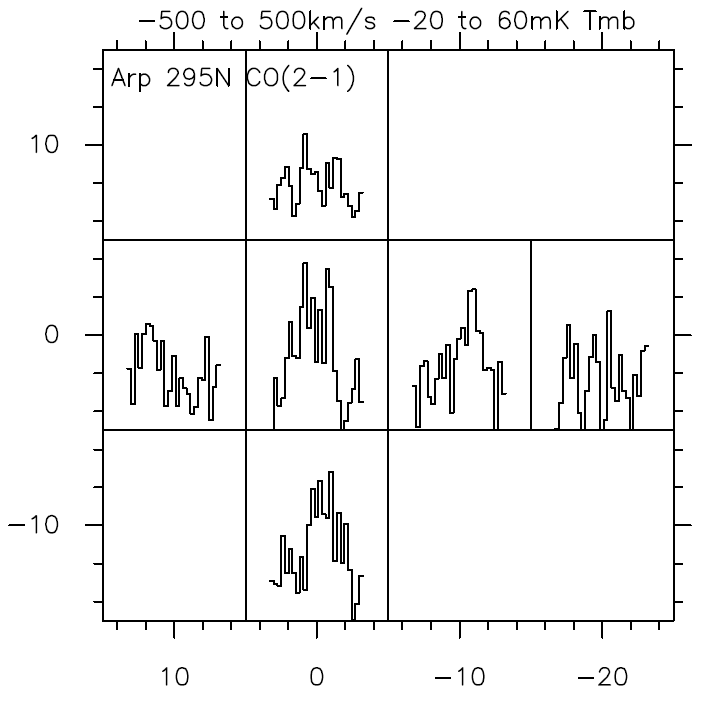}
}
\caption{CO(2-1) map of spectra in Arp~295N.
}
\label{fig:mapa295n_2}
\end{figure}

The northern galaxy of the Arp~295 pair is a nearly edge-on spiral,
which is strongly perturbed by the interaction, with disordered
shells and ripples. The observed spectra of Fig. \ref{fig:mapa295n}
and \ref{fig:mapa295n_2} are essentially compatible with molecular gas 
in the disk, with some extension in the western shell (see Table \ref{tab:CO-results}). However,
the emission does not extend far in the most western offset, so it is
still marginal.

\section{Discussion}
\label{disc}

 The dynamical mechanism to form shells through phase-wrapping has been described in detail
\citep{Quinn1984,Dupraz1986,Dupraz1987}. Stars from a companion, liberated by tidal forces,
oscillate in the primary galaxy potential, and a sharp edge forms as an outward propagating
wave, at their apocenters. There could be radial oscillation in a prolate potential, forming
alternate shells with zero velocity at their edge, or less radial ones, with angular momentum,
forming shells around the equatorial plane of an oblate potential, keeping some rotational
velocity at their edge. Another mechanism is space-wrapping, where shells occur as ordinary
loops in interactions and mergers, with larger velocities.

\subsection{Molecular clouds in shells}
Since gas is dissipative, it may appear surprising to find gas in shells.  During
radial oscillations, the diffuse gas should fall into the center.
Dense gas clouds do experience much less dissipation than diffuse HI gas. Moreover, 
numerical modeling of the formation of shells \citep{Combes2000},
using a realistic distribution of its stellar and gaseous component, showed that in fact one can easily
reproduce the observed spatial distribution of gaseous and stellar shells within the standard models
of the theory of shell formation. A confirmation of this model was the detection of molecular gas 
associated with the shells in Centaurus A  \citep{Charmandaris2000}. Using SEST
telescope, they detected a molecular gas of $4.3 \times 10^{7}$ M$_{\odot}$ with the same kinematical 
characteristics as the HI shells.

The model predicts that a significant fraction of the merger-remnant dense molecular clouds should 
have similar dynamical behavior to the stellar component. The present detection of molecular clouds 
in shells through CO emission supports the model. With higher spatial resolution, it will be
possible to determine the difference in position between the gas and stars in shells, and infer the
loss of angular momentum during the encounter.

\subsection{CO velocity in shells}

CO emission in shells were clearly detected for three galaxies: Arp~10, NGC~3656 and
NGC~3934. In the case of Arp~10, the inclination on the sky plane is small
 (i=14$^\circ$ from Hyperleda), and the
small velocity difference with respect to the systemic 
recession velocity for the gas in the shells is compatible with both phase and space
wrapping. For the two other galaxies, the inclination is relatively large
(i=51$^\circ$ and i=46$^\circ$ for NGC~3656 and 3934 respectively), and 
the average velocity with respect to the systemic velocity of the shells is 
relatively small. However, the CO
profiles in the shells are broad, and even in both NGC~3656 and NGC~3934 show a two-horn shape.
This means that the CO beam is covering a large shell region, where the material
is not only in radial orbits, but keeps a tangential velocity at apocenter. In the line of
sight towards the observer, are projected both some material with positive and
negative angular momentum with respect to the galaxy.

This configuration is impossible
to encounter with a disk in rotation. Two-horn profiles are only seen when the observation
beam is covering a wide region near the center, including both the approaching and receding sides.
For an observation far from the center of the rotating disk, only one velocity peak (either blue or red)
is expected.
For shells on the contrary, the material from the companion is infalling on a nearly radial orbit,
and there may exist part of the matter arriving with a positive, and part with a negative
angular momentum with respect to the primary galaxy. In the shell region far from the center,
it is possible to have the two velocity signs towards the same line of sight.
The two-horn profile shape is therefore supporting the CO detection in shells.  Confirmation
of this scenario is required, through observations with higher spatial resolution.

\subsection{Comparison with HI in shells}
HI has been mapped in Arp~230 \citep{Schiminovich2013}, NGC~474 \citep{Rampazzo2006},
Arp~10 \citep{Charmandaris1996}, NGC~3032 \citep{Oosterloo2010},
NGC~3656 \citep{Balcells2001}, Arp~295 \citep{Hibbard1996}.
In all these 6 galaxies, the HI is extended enough to cover the shells,
and in NGC~3656, the HI is clearly seen associated with the southern 
stellar shell, with compatible kinematics. In most cases, however, the spatial 
resolution is not sufficient to clearly conclude that the gas provided
by a small merging companion has been crossing the system, and oscillates
in the shells. The only system where the presence of HI gas in shells
is clear is in NGC~3656, that \cite{Balcells2001} compare to NGC~5128 (Centaurus A),
where the first HI shells were discovered \citep{Schiminovich1994}.
Given the orientation of the gas and dusty disk, perpendicular
to the main stellar body, this shell system must have been formed through
phase-wrapping of material in radial orbits. The presence of HI and CO emission
in the shells can only be explained by the presence of a clumpy low-filling
factor gas, with low dissipation \citep{Combes2000}.

\section{Summary}
\label{conclu}
We have carried out CO(1-0) and CO(2-1) line observations of nine shell galaxies with the
IRAM-30m telescope. Molecular gas was detected in six of them, together with two of the companions.
In three of the detected galaxies (Arp~230, NGC~3032 and Arp~205N), we cannot conclude
that gas is detected in the shells,
the detection is compatible with gas in the inner rotating disk.
In the three remaining galaxies (Arp~10, NGC~3656 and NGC~3934), molecular gas is clearly detected
in the shells, with a velocity close to systemic, and not in continuity with the inner
rotating disk. The amount of molecular gas is estimated at M(H$_2$) = 3.4 10$^9$ M$_\odot$
in the shells outside the ring in Arp~10, 3.8 10$^8$ M$_\odot$ in the southern shell of
NGC~3656, and 2.6 10$^8$ M$_\odot$ in the two detected shells of NGC~3934.

The mechanism to form those gaseous shells might be different. For two cases at least (Arp~10, NGC~3656),
the encounter with the small companion had a very small impact parameter, and the phase-wrapping
in a nearly radial orbit must be playing the major role. The morphology of Arp~10 corresponds
to an oblate system, as well as most shells in NGC~3656. However, the major shell might be only
the end of the tidal tail.
The gas of the companion of NGC~3656 has left a radial trace, visible as a contrasted dust
lane, and the major shell is found in continuity, indicating the companion orbit, 
and that of the tidal debris.
NGC~3934 main body is seen nearly edge-on on the sky, and the situation is less clear. 
Shells distributed
symmetrically at about 40$^\circ$ of the minor axis are clearly detected in molecular gas.
The fact that gas can be detected in shells tend to support the clumpiness and low dissipative
character of a large part of the interstellar medium. 
 The line profiles in the shells reveal a two-horn shape, indicating the
presence of both positive and negative angular momentum relative to the primary galaxy.
This supports the CO detection in shells, since it is impossible to explain a two-horn
profile in the outer parts of a rotating galaxy disk. Material from the companion
must have infallen in a nearly radial orbit, part of it with a positive and the other
part a negative impact parameter. The material acquired a tangential velocity at apocenter.
The shells are then also phase-wrapped. Follow-up with higher spatial resolution
is required to test the relative position of gas and stellar shells, to confirm the dynamical mechanism.

\begin{acknowledgements}
 We warmly thank the referee for constructive comments and suggestions. 
 This work is based on observations carried out with the IRAM 30m telescope. IRAM is supported by INSU/CNRS (France), MPG (Germany) and IGN (Spain).
This publication has made use of data products from the NASA/IPAC Extragalactic Database (NED).
 We acknowledge the usage of the HyperLeda database (http://leda.univ-lyon1.fr).
\end{acknowledgements}

\bibliographystyle{aa}
\bibliography{shells}

\begin{thebibliography}{60}
\expandafter\ifx\csname natexlab\endcsname\relax\def\natexlab#1{#1}\fi

\bibitem[{{Alatalo} {et~al.}(2013){Alatalo}, {Davis}, {Bureau}, {Young},
  {Blitz}, {Crocker}, {Bayet}, {Bois}, {Bournaud}, {Cappellari}, {Davies}, {de
  Zeeuw}, {Duc}, {Emsellem}, {Khochfar}, {Krajnovi{\'c}}, {Kuntschner},
  {Lablanche}, {Morganti}, {McDermid}, {Naab}, {Oosterloo}, {Sarzi}, {Scott},
  {Serra}, \& {Weijmans}}]{Alatalo2013}
{Alatalo}, K., {Davis}, T.~A., {Bureau}, M., {et~al.} 2013, \mnras, 432, 1796

\bibitem[{{Appleton} \& {Struck-Marcell}(1996)}]{Appleton1996}
{Appleton}, P.~N. \& {Struck-Marcell}, C. 1996, \fcp, 16, 111

\bibitem[{{Arp}(1966)}]{Arp1966}
{Arp}, H. 1966, \apjs, 14, 1

\bibitem[{{Balcells} \& {Stanford}(1990)}]{Balcells1990}
{Balcells}, M. \& {Stanford}, S.~A. 1990, \apj, 362, 443

\bibitem[{{Balcells} {et~al.}(2001){Balcells}, {van Gorkom}, {Sancisi}, \& {del
  Burgo}}]{Balcells2001}
{Balcells}, M., {van Gorkom}, J.~H., {Sancisi}, R., \& {del Burgo}, C. 2001,
  \aj, 122, 1758

\bibitem[{{Barnes} \& {Hernquist}(1992)}]{Barnes1992}
{Barnes}, J.~E. \& {Hernquist}, L. 1992, \araa, 30, 705

\bibitem[{{Bettoni} {et~al.}(2011){Bettoni}, {Galletta}, {Rampazzo}, {Marino},
  {Mazzei}, \& {Buson}}]{Bettoni2011}
{Bettoni}, D., {Galletta}, G., {Rampazzo}, R., {et~al.} 2011, \aap, 534, A24

\bibitem[{{Bizyaev} {et~al.}(2007){Bizyaev}, {Moiseev}, \&
  {Vorobyov}}]{Bizyaev2007}
{Bizyaev}, D.~V., {Moiseev}, A.~V., \& {Vorobyov}, E.~I. 2007, \apj, 662, 304

\bibitem[{{Bolatto} {et~al.}(2013){Bolatto}, {Wolfire}, \&
  {Leroy}}]{Bolatto2013}
{Bolatto}, A.~D., {Wolfire}, M., \& {Leroy}, A.~K. 2013, \araa, 51, 207

\bibitem[{{Buson} {et~al.}(2004){Buson}, {Bertola}, {Bressan}, {Burstein}, \&
  {Cappellari}}]{Buson2004}
{Buson}, L.~M., {Bertola}, F., {Bressan}, A., {Burstein}, D., \& {Cappellari},
  M. 2004, \aap, 423, 965

\bibitem[{{Carter} {et~al.}(1982){Carter}, {Allen}, \& {Malin}}]{Carter1982}
{Carter}, D., {Allen}, D.~A., \& {Malin}, D.~F. 1982, \nat, 295, 126

\bibitem[{{Charmandaris} \& {Appleton}(1996)}]{Charmandaris1996}
{Charmandaris}, V. \& {Appleton}, P.~N. 1996, \apj, 460, 686

\bibitem[{{Charmandaris} {et~al.}(2000){Charmandaris}, {Combes}, \& {van der
  Hulst}}]{Charmandaris2000}
{Charmandaris}, V., {Combes}, F., \& {van der Hulst}, J.~M. 2000, \aap, 356, L1

\bibitem[{{Chien} {et~al.}(2007){Chien}, {Barnes}, {Kewley}, \&
  {Chambers}}]{Chien2007}
{Chien}, L.-H., {Barnes}, J.~E., {Kewley}, L.~J., \& {Chambers}, K.~C. 2007,
  \apjl, 660, L105

\bibitem[{{Combes} \& {Charmandaris}(2000)}]{Combes2000}
{Combes}, F. \& {Charmandaris}, V. 2000, in Astronomical Society of the Pacific
  Conference Series, Vol. 197, Dynamics of Galaxies: from the Early Universe to
  the Present, ed. F.~{Combes}, G.~A. {Mamon}, \& V.~{Charmandaris}, 339

\bibitem[{{Combes} {et~al.}(2007){Combes}, {Young}, \& {Bureau}}]{Combes2007}
{Combes}, F., {Young}, L.~M., \& {Bureau}, M. 2007, \mnras, 377, 1795

\bibitem[{{Cooper} {et~al.}(2011){Cooper}, {Mart{\'{\i}}nez-Delgado}, {Helly},
  {Frenk}, {Cole}, {Crawford}, {Zibetti}, {Carballo-Bello}, \&
  {GaBany}}]{Cooper2011}
{Cooper}, A.~P., {Mart{\'{\i}}nez-Delgado}, D., {Helly}, J., {et~al.} 2011,
  \apjl, 743, L21

\bibitem[{{Duc} {et~al.}(2015){Duc}, {Cuillandre}, {Karabal}, {Cappellari},
  {Alatalo}, {Blitz}, {Bournaud}, {Bureau}, {Crocker}, {Davies}, {Davis}, {de
  Zeeuw}, {Emsellem}, {Khochfar}, {Krajnovi{\'c}}, {Kuntschner}, {McDermid},
  {Michel-Dansac}, {Morganti}, {Naab}, {Oosterloo}, {Paudel}, {Sarzi}, {Scott},
  {Serra}, {Weijmans}, \& {Young}}]{Duc2015}
{Duc}, P.-A., {Cuillandre}, J.-C., {Karabal}, E., {et~al.} 2015, \mnras, 446,
  120

\bibitem[{{Dupraz} \& {Combes}(1986)}]{Dupraz1986}
{Dupraz}, C. \& {Combes}, F. 1986, \aap, 166, 53

\bibitem[{{Dupraz} \& {Combes}(1987)}]{Dupraz1987}
{Dupraz}, C. \& {Combes}, F. 1987, \aap, 185, L1

\bibitem[{{Galletta} {et~al.}(1997){Galletta}, {Sage}, \&
  {Sparke}}]{Galletta1997}
{Galletta}, G., {Sage}, L.~J., \& {Sparke}, L.~S. 1997, \mnras, 284, 773

\bibitem[{{Hernquist} \& {Quinn}(1988)}]{Hernquist1988}
{Hernquist}, L. \& {Quinn}, P.~J. 1988, \apj, 331, 682

\bibitem[{{Hernquist} \& {Quinn}(1989)}]{Hernquist1989}
{Hernquist}, L. \& {Quinn}, P.~J. 1989, \apj, 342, 1

\bibitem[{{Hibbard} \& {van Gorkom}(1996)}]{Hibbard1996}
{Hibbard}, J.~E. \& {van Gorkom}, J.~H. 1996, \aj, 111, 655

\bibitem[{{Huchtmeier}(1982)}]{Huchtmeier1982}
{Huchtmeier}, W.~K. 1982, \aap, 110, 121

\bibitem[{{Kim} {et~al.}(1988){Kim}, {Guhathakurta}, {van Gorkom}, {Jura}, \&
  {Knapp}}]{Kim1988}
{Kim}, D.-W., {Guhathakurta}, P., {van Gorkom}, J.~H., {Jura}, M., \& {Knapp},
  G.~R. 1988, \apj, 330, 684

\bibitem[{{Lim} {et~al.}(2017){Lim}, {Peng}, {Duc}, {Fensch}, {Durrell},
  {Harris}, {Cuillandre}, {Gwyn}, {Lan{\c c}on}, \&
  {S{\'a}nchez-Janssen}}]{Lim2017}
{Lim}, S., {Peng}, E.~W., {Duc}, P.-A., {et~al.} 2017, \apj, 835, 123

\bibitem[{{Lucero} \& {Young}(2013)}]{Lucero2013}
{Lucero}, D.~M. \& {Young}, L.~M. 2013, \aj, 145, 56

\bibitem[{{McDermid} {et~al.}(2006){McDermid}, {Emsellem}, {Shapiro}, {Bacon},
  {Bureau}, {Cappellari}, {Davies}, {de Zeeuw}, {Falc{\'o}n-Barroso},
  {Krajnovi{\'c}}, {Kuntschner}, {Peletier}, \& {Sarzi}}]{McDermid2006}
{McDermid}, R.~M., {Emsellem}, E., {Shapiro}, K.~L., {et~al.} 2006, \mnras,
  373, 906

\bibitem[{{McGaugh} \& {Bothun}(1990)}]{McGaugh1990}
{McGaugh}, S.~S. \& {Bothun}, G.~D. 1990, \aj, 100, 1073

\bibitem[{{Oosterloo} {et~al.}(2010){Oosterloo}, {Morganti}, {Crocker},
  {J{\"u}tte}, {Cappellari}, {de Zeeuw}, {Krajnovi{\'c}}, {McDermid},
  {Kuntschner}, {Sarzi}, \& {Weijmans}}]{Oosterloo2010}
{Oosterloo}, T., {Morganti}, R., {Crocker}, A., {et~al.} 2010, \mnras, 409, 500

\bibitem[{{Prieur}(1988)}]{Prieur1988}
{Prieur}, J.-L. 1988, \apj, 326, 596

\bibitem[{{Quinn}(1984)}]{Quinn1984}
{Quinn}, P.~J. 1984, \apj, 279, 596

\bibitem[{{Rampazzo} {et~al.}(2006){Rampazzo}, {Alexander}, {Carignan},
  {Clemens}, {Cullen}, {Garrido}, {Marcelin}, {Sheth}, \&
  {Trinchieri}}]{Rampazzo2006}
{Rampazzo}, R., {Alexander}, P., {Carignan}, C., {et~al.} 2006, \mnras, 368,
  851

\bibitem[{{Rampazzo} {et~al.}(2007){Rampazzo}, {Marino}, {Tantalo}, {Bettoni},
  {Buson}, {Chiosi}, {Galletta}, {Gr{\"u}tzbauch}, \& {Rich}}]{Rampazzo2007}
{Rampazzo}, R., {Marino}, A., {Tantalo}, R., {et~al.} 2007, \mnras, 381, 245

\bibitem[{{Renaud} {et~al.}(2015){Renaud}, {Bournaud}, \& {Duc}}]{Renaud2015}
{Renaud}, F., {Bournaud}, F., \& {Duc}, P.-A. 2015, \mnras, 446, 2038

\bibitem[{{Richter} {et~al.}(1994){Richter}, {Sackett}, \&
  {Sparke}}]{Richter1994}
{Richter}, O.-G., {Sackett}, P.~D., \& {Sparke}, L.~S. 1994, \aj, 107, 99

\bibitem[{{Sage} \& {Wrobel}(1989)}]{Sage1989}
{Sage}, L.~J. \& {Wrobel}, J.~M. 1989, \apj, 344, 204

\bibitem[{{Sansom} {et~al.}(2000){Sansom}, {Hibbard}, \&
  {Schweizer}}]{Sansom2000}
{Sansom}, A.~E., {Hibbard}, J.~E., \& {Schweizer}, F. 2000, \aj, 120, 1946

\bibitem[{{Schiminovich} {et~al.}(1997){Schiminovich}, {van Gorkom}, {van der
  Hulst}, {Oosterloo}, \& {Wilkinson}}]{Schiminovich1997}
{Schiminovich}, D., {van Gorkom}, J., {van der Hulst}, T., {Oosterloo}, T., \&
  {Wilkinson}, A. 1997, in Astronomical Society of the Pacific Conference
  Series, Vol. 116, The Nature of Elliptical Galaxies; 2nd Stromlo Symposium,
  ed. M.~{Arnaboldi}, G.~S. {Da Costa}, \& P.~{Saha}, 362

\bibitem[{{Schiminovich} {et~al.}(2013){Schiminovich}, {van Gorkom}, \& {van
  der Hulst}}]{Schiminovich2013}
{Schiminovich}, D., {van Gorkom}, J.~H., \& {van der Hulst}, J.~M. 2013, \aj,
  145, 34

\bibitem[{{Schiminovich} {et~al.}(1994){Schiminovich}, {van Gorkom}, {van der
  Hulst}, \& {Kasow}}]{Schiminovich1994}
{Schiminovich}, D., {van Gorkom}, J.~H., {van der Hulst}, J.~M., \& {Kasow}, S.
  1994, \apjl, 423, L101

\bibitem[{{Schiminovich} {et~al.}(1995){Schiminovich}, {van Gorkom}, {van der
  Hulst}, \& {Malin}}]{Schiminovich1995}
{Schiminovich}, D., {van Gorkom}, J.~H., {van der Hulst}, J.~M., \& {Malin},
  D.~F. 1995, \apjl, 444, L77

\bibitem[{{Schweizer}(1990)}]{Schweizer1990}
{Schweizer}, F. 1990, {Interactions in our time.}, ed. R.~{Wielen}, 60--71

\bibitem[{{Schweizer} \& {Seitzer}(1992)}]{Schweizer1992}
{Schweizer}, F. \& {Seitzer}, P. 1992, \aj, 104, 1039

\bibitem[{{Serra} {et~al.}(2008){Serra}, {Trager}, {Oosterloo}, \&
  {Morganti}}]{Serra2008}
{Serra}, P., {Trager}, S.~C., {Oosterloo}, T.~A., \& {Morganti}, R. 2008, \aap,
  483, 57

\bibitem[{{Sofue} {et~al.}(1993){Sofue}, {Wakamatsu}, {Taniguchi}, \&
  {Nakai}}]{Sofue1993}
{Sofue}, Y., {Wakamatsu}, K.-I., {Taniguchi}, Y., \& {Nakai}, N. 1993, \pasj,
  45, 43

\bibitem[{{Solomon} \& {Vanden Bout}(2005)}]{Solomon2005}
{Solomon}, P.~M. \& {Vanden Bout}, P.~A. 2005, \araa, 43, 677

\bibitem[{{Tateuchi} {et~al.}(2015){Tateuchi}, {Konishi}, {Motohara},
  {Takahashi}, {Mitani Kato}, {Kitagawa}, {Todo}, {Toshikawa}, {Sako},
  {Uchimoto}, {Ohsawa}, {Asano}, {Ita}, {Kamizuka}, {Komugi}, {Koshida},
  {Manabe}, {Nakamura}, {Nakashima}, {Okada}, {Takagi}, {Tanab{\'e}},
  {Uchiyama}, {Aoki}, {Doi}, {Handa}, {Kawara}, {Kohno}, {Minezaki}, {Miyata},
  {Morokuma}, {Soyano}, {Tamura}, {Tanaka}, {Tarusawa}, \&
  {Yoshii}}]{Tateuchi2015}
{Tateuchi}, K., {Konishi}, M., {Motohara}, K., {et~al.} 2015, \apjs, 217, 1

\bibitem[{{Toomre} \& {Toomre}(1972)}]{Toomre1972}
{Toomre}, A. \& {Toomre}, J. 1972, \apj, 178, 623

\bibitem[{{Turnbull} {et~al.}(1999){Turnbull}, {Bridges}, \&
  {Carter}}]{Turnbull1999}
{Turnbull}, A.~J., {Bridges}, T.~J., \& {Carter}, D. 1999, \mnras, 307, 967

\bibitem[{{Ueda} {et~al.}(2014){Ueda}, {Iono}, {Yun}, {Crocker}, {Narayanan},
  {Komugi}, {Espada}, {Hatsukade}, {Kaneko}, {Matsuda}, {Tamura}, {Wilner},
  {Kawabe}, \& {Pan}}]{Ueda2014}
{Ueda}, J., {Iono}, D., {Yun}, M.~S., {et~al.} 2014, \apjs, 214, 1

\bibitem[{{van Driel} {et~al.}(2000){van Driel}, {Arnaboldi}, {Combes}, \&
  {Sparke}}]{vanDriel2000}
{van Driel}, W., {Arnaboldi}, M., {Combes}, F., \& {Sparke}, L.~S. 2000, \aaps,
  141, 385

\bibitem[{{van Gorkom}(2000)}]{vanGorkom2000}
{van Gorkom}, J. 2000, in Radio interferometry : the saga and the science, ed.
  D.~G. {Finley} \& W.~M. {Goss}, 137

\bibitem[{{Weil} \& {Hernquist}(1993)}]{Weil1993}
{Weil}, M.~L. \& {Hernquist}, L. 1993, \apj, 405, 142

\bibitem[{{Whitmore} {et~al.}(1990){Whitmore}, {Lucas}, {McElroy},
  {Steiman-Cameron}, {Sackett}, \& {Olling}}]{Whitmore1990}
{Whitmore}, B.~C., {Lucas}, R.~A., {McElroy}, D.~B., {et~al.} 1990, \aj, 100,
  1489

\bibitem[{{Whitmore} {et~al.}(1993){Whitmore}, {Schweizer}, {Leitherer},
  {Borne}, \& {Robert}}]{Whitmore1993}
{Whitmore}, B.~C., {Schweizer}, F., {Leitherer}, C., {Borne}, K., \& {Robert},
  C. 1993, \aj, 106, 1354

\bibitem[{{Wiklind} {et~al.}(1995){Wiklind}, {Combes}, \&
  {Henkel}}]{Wiklind1995}
{Wiklind}, T., {Combes}, F., \& {Henkel}, C. 1995, \aap, 297, 643

\bibitem[{{Young}(2002)}]{Young2002}
{Young}, L.~M. 2002, \aj, 124, 788

\bibitem[{{Young} {et~al.}(2011){Young}, {Bureau}, {Davis}, {Combes},
  {McDermid}, {Alatalo}, {Blitz}, {Bois}, {Bournaud}, {Cappellari}, {Davies},
  {de Zeeuw}, {Emsellem}, {Khochfar}, {Krajnovi{\'c}}, {Kuntschner},
  {Lablanche}, {Morganti}, {Naab}, {Oosterloo}, {Sarzi}, {Scott}, {Serra}, \&
  {Weijmans}}]{Young2011}
{Young}, L.~M., {Bureau}, M., {Davis}, T.~A., {et~al.} 2011, \mnras, 414, 940

\end{thebibliography}

\appendix

\section{The molecular content of companions}

\begin{figure}[h!]
\centerline{
\includegraphics[angle=0,width=8cm]{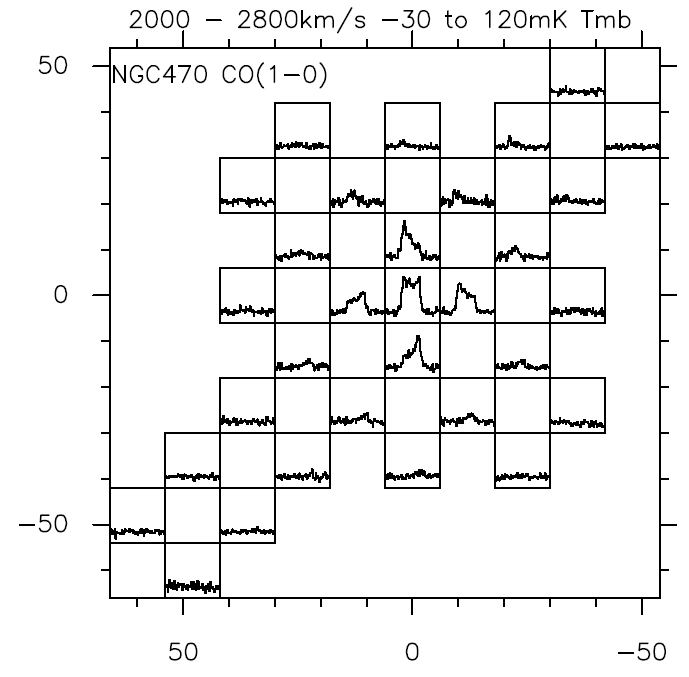}
}
\caption{CO(1-0) map of spectra in NGC~470.
}
\label{fig:map470}
\end{figure}

\begin{figure}[h!]
\centerline{
\includegraphics[angle=0,width=8cm]{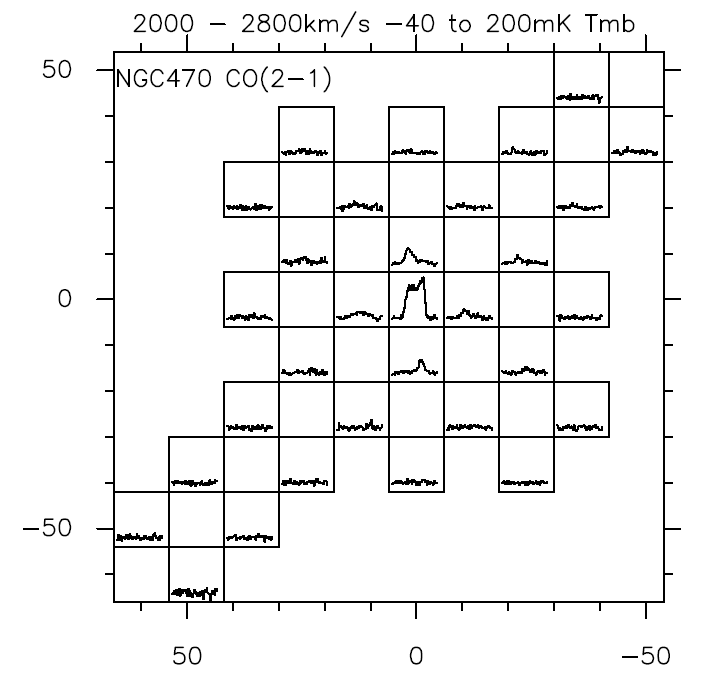}
}
\caption{CO(2-1) map of spectra in NGC~470.
}
\label{fig:map470_21}
\end{figure}

\begin{figure}[h!]
\centerline{
\includegraphics[angle=0,width=8cm]{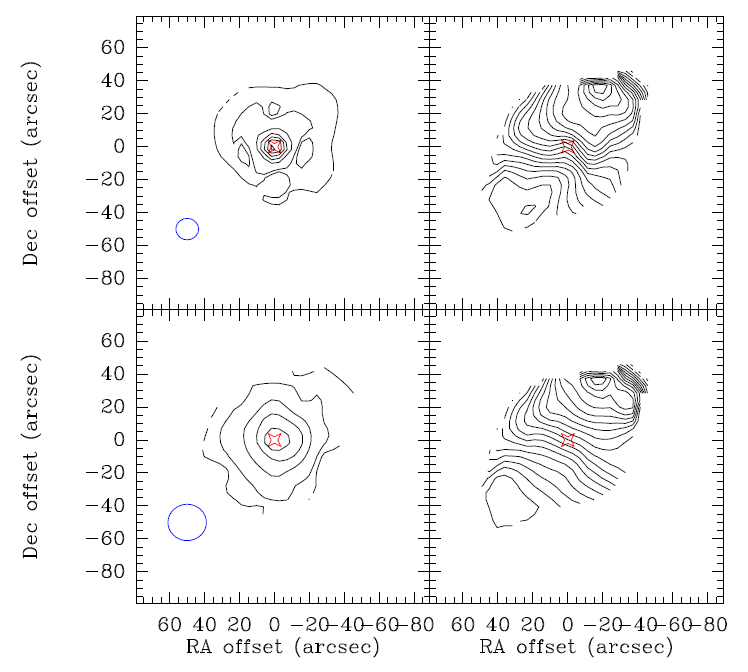}
}
\caption{Isocontours of NGC~470 for CO(2-1) (Upper panels) and CO(1-0) (Lower panels),
	the right panels display the corresponding velocity fields. The isovels are distant
	by 20 \kms, they range from 2200 \kms (NW) to 2560 \kms (SE).
}
\label{fig:carte470}
\end{figure}

The companion of the shell galaxy NGC~474 (Arp~227) is a late-type
spiral galaxy, NGC~470, which is rich in HI and CO emission. 
 It is likely that the HI extension covering NGC~474 (see Fig. \ref{fig:HI-474})
 has been accreted from this companion. The beams observed on
  NGC~470 are displayed in Fig. \ref{fig:n474-beams}. The spectra maps
in CO(1-0) and CO(2-1) are displayed in Fig. \ref{fig:map470} and
\ref{fig:map470_21} respectively. Isocontours of the surface density
and the velocity fields are plotted in \ref{fig:carte470}.

\begin{figure}[h!]
\centerline{
\includegraphics[angle=0,width=7.5cm]{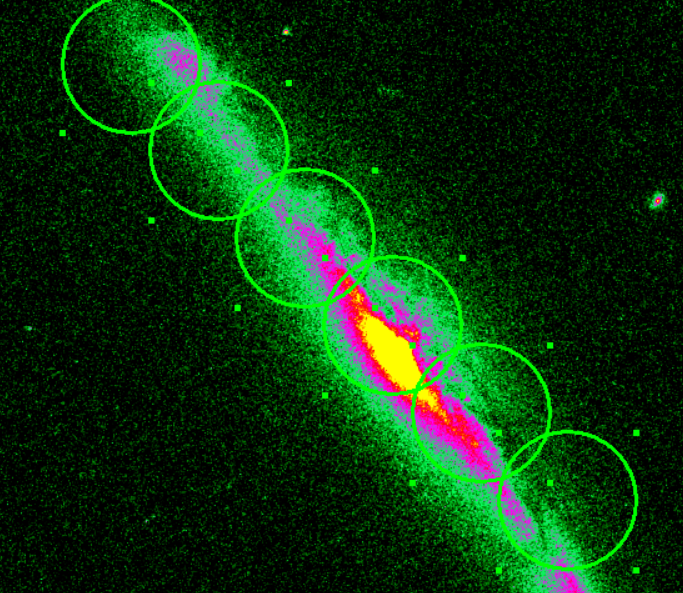}
}
	\caption{CO(2-1) 11'' beams observed
	towards Arp~295S, overlaid on an HST  F814W image.
}
\label{a295s-beams}
\end{figure}

\begin{figure}[h!]
\centerline{
\includegraphics[angle=0,width=8cm]{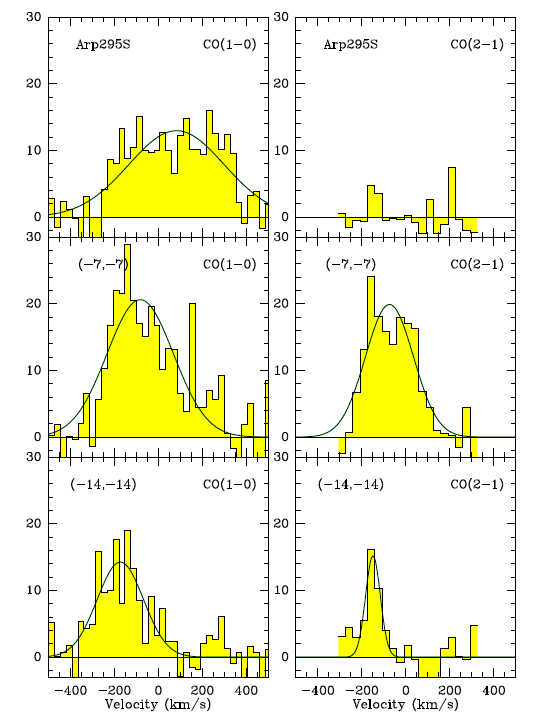}
}
\caption{CO spectra in Arp~295S, central position and offsets 
(in arcsec). The vertical axis is main beam temperature in mK. }
\label{fig-a295s}
\end{figure}

\begin{figure}[h!]
\centerline{
\includegraphics[angle=0,width=8cm]{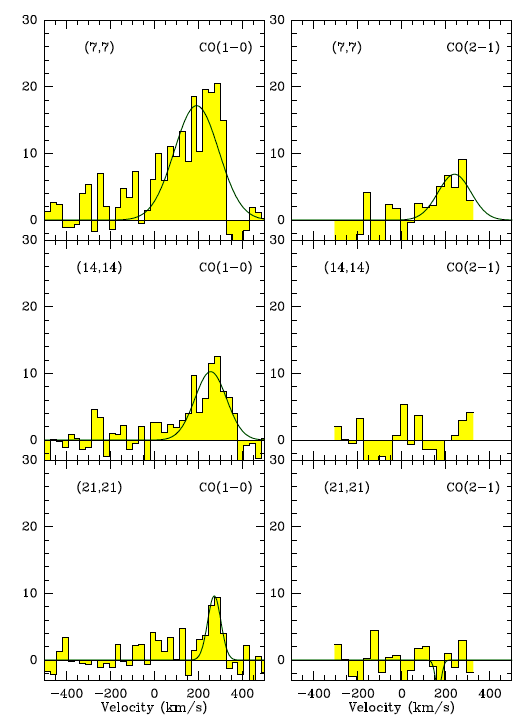}
}
	\caption{CO spectra in Arp~295S (following), northern offsets 
(in arcsec). The vertical axis is main beam temperature in mK. }
\label{fig-a295sb}
\end{figure}

The other member of the pair in Arp~295 is an edge-on spiral galaxy,
rich in molecular gas. The pointings are displayed on an HST F814W image
in Fig. \ref{a295s-beams}.
CO spectra in Arp~295S are plotted in
Fig. \ref{fig-a295s} and \ref{fig-a295sb} . The molecular content
in each beam is detailed in Table \ref{tab:companions}.

\begin{table*}[htb]
\caption{CO results from the companions}
\centering
\label{tab:companions}
\begin{tabular}{rccccccccc}
\noalign{\smallskip} \hline \hline \noalign{\smallskip}
        &  &  &  CO(1-0) &  & &  CO(2-1) & & &\\
Galaxy & Offsets & Int-Flux & V &$\Delta$V & Int-Flux  & V  & $\Delta$V & M(H$_2$) &$\Sigma$(H$_2$)\\
& ('','')   & Jy\kms & \kms & \kms & Jy\kms & \kms & \kms  & $10^8$ M$_\odot$&  M$_\odot$pc$^{-2}$\\
 \hline
NGC~470&(0,0)&126$\pm$2&66$\pm$2&252$\pm$4&216$\pm$2&88$\pm$1&246$\pm$2&14$\pm$0.2&151\\
      & see &31 pt& map&  & & & & &\\
 \hline
ARP~295S&(0,0)&35.3$\pm$3 &82$\pm$22&512$\pm$48& $<$4 & -- & -- &32.5$\pm$3& 42\\
& (-7,-7)& 39$\pm$3.5&-82$\pm$16&355$\pm$40&27.1$\pm$2.5&-74$\pm$11&256$\pm$24&36$\pm$3&47\\
& (-14,-14)&19$\pm$1.6&-173$\pm$10&248$\pm$25&6.4$\pm$1.3&-150$\pm$7&80$\pm$23&17$\pm$1.5&22\\
& (7,7)& 22$\pm$2&-192$\pm$12&238$\pm$27&6.5$\pm$3&239$\pm$38&176$\pm$96&20.3$\pm$2&26\\
& (14,14)&9.2$\pm$1.1&256$\pm$11&169$\pm$25&$<$2.5& --& -- &8.5$\pm$1&11\\
& (21,21)&3.6$\pm$0.7&272$\pm$7&71$\pm$17& $<$2.5& --& -- &3.3$\pm$0.6&4\\
\noalign{\smallskip} \hline \noalign{\smallskip}
\end{tabular}
\begin{list}{}{}
\item --Integrated flux, velocity and FWHM obtained through gaussian fits 
      -- Velocities are counted from the central redshifts indicated
	in Table \ref{tab:param} for the companion galaxy
      -- The H$_2$ masses are estimated within the CO(1-0) beam of 23'', with
	the standard conversion factor of the Milky Way $\alpha$ = 
	4.36~M$_\odot~({\rm K}~{\rm km}~{\rm s}^{-1}~{\rm pc}^{2})^{-1}$.
       -- Upper limits are given at 3$\sigma$, assuming a FWHM of 300\kms.
\end{list}
\end{table*}

\end{document}